\documentclass[preprint,12pt]{aastex}
\usepackage{latexsym,times,epsf}

\def\spose#1{\hbox to 0pt{#1\hss}}

\def\lya{\ifmmode {\rm\,Ly\alpha}\else ${\rm\,Ly\alpha}$\fi}
\def\Mdot {\ifmmode {\rm {\dot M}} \else ${\rm {\dot M}}$\fi}

\def\kms{\ifmmode {\rm\,km\,s^{-1}}\else
    ${\rm\,km\,s^{-1}}$\fi}
\def\kmsMpc{\ifmmode {\rm\,km\,s^{-1}\,Mpc^{-1}}\else
    ${\rm\,km\,s^{-1}\,Mpc^{-1}}$\fi}

\def\msun{\ifmmode {\rm\,M_\odot}\else ${\rm\,M_\odot}$\fi}
\def\Msun{\ifmmode {\rm\,M_\odot}\else ${\rm\,M_\odot}$\fi}
\def\lsun{\ifmmode {\rm\,L_\odot}\else ${\rm\,L_\odot}$\fi}
\def\Lsun{\ifmmode {\rm\,L_\odot}\else ${\rm\,L_\odot}$\fi}
\def\rsun{\ifmmode {\rm\,R_\odot}\else ${\rm\,R_\odot}$\fi}
\def\Rsun{\ifmmode {\rm\,R_\odot}\else ${\rm\,R_\odot}$\fi}

\def\cm{{\rm\,cm}}
\def\cm3{\ifmmode {\rm\,cm^{-3}}\else ${\rm\,cm^{-3}}$\fi}

\def\ergps{\ifmmode {\rm\,erg\,s^{-1}}\else ${\rm\,erg\,s^{-1}}$\fi}
\def\ergpscm2{\ifmmode {\rm\,erg\,s^{-1}\,cm^{-2}}\else
    ${\rm\,erg\,s^{-1}\,cm^{-2}}$\fi}

\def\deg{\ifmmode {^{\circ}}\else {$^\circ$}\fi}
\def\degr{\ifmmode {^{\circ}}\else {$^\circ$}\fi}
\def\degs{\ifmmode {^{\circ}}\else {$^\circ$}\fi}

\def\h3Mpc{h^{-3}{\rm Mpc}^3}
\def\Ho{\ifmmode {\rm\,H_0}\else ${\rm\,H_0}$\fi}
\def\hnot{\ifmmode {\rm\,H_0}\else ${\rm\,H_0}$\fi}
\def\h0{\ifmmode {\rm\,H_0}\else ${\rm\,H_0}$\fi}
\def\hnotunit{\ifmmode {\rm\,km\,s^{-1}\,Mpc^{-1}}\else
    ${\rm\,km\,s^{-1}\,Mpc^{-1}}$\fi}
\def\qnot{\ifmmode {\rm\,q_0}\else ${\rm q_0}$\fi}
\def\q0{\ifmmode {\rm\,q_0}\else ${\rm q_0}$\fi}

\def\mic{\ifmmode {\rm\,\mu m}\else ${\rm \mu m}$\fi}
\def\micron{\ifmmode {\rm\,\mu m}\else ${\rm \mu m}$\fi}
\def\microns{\ifmmode {\rm\,\mu m}\else ${\rm \mu m}$\fi}

\def\arcsec{\ifmmode {^{\prime\prime}}\else $^{\prime\prime}$\fi}
\def\asec{\ifmmode {^{\prime\prime}}\else $^{\prime\prime}$\fi}
\def\arcmin{\ifmmode {^{\prime}}\else $^{\prime}$\fi}
\def\amin{\ifmmode {^{\prime}}\else $^{\prime}$\fi}

\def\secper{\ifmmode \rlap.{^{s}}\else $\rlap{.}{^{s}} $\fi}
\def\minper{\ifmmode \rlap.{^{m}}\else $\rlap{.}{^m} $\fi}
\def\magper{\ifmmode \rlap.{^{m}}\else $\rlap{.}{^m} $\fi}
\def\farcs{\ifmmode \rlap.{^{\prime\prime}}\else
    $\rlap.{^{\prime\prime}}$\fi}
\def\arcsper{\ifmmode \rlap.{^{\prime\prime}}\else
    $\rlap.{^{\prime\prime}}$\fi}
\def\arcmper{\ifmmode \rlap.{^{\prime}}\else
    $\rlap.{^{\prime}}$\fi}
\def\spose#1{\hbox to 0pt{#1\hss}}
\def\simlt{\mathrel{\spose{\lower 3pt\hbox{$\mathchar"218$}}
     \raise 2.0pt\hbox{$\mathchar"13C$}}}
\def\simgt{\mathrel{\spose{\lower 3pt\hbox{$\mathchar"218$}}
     \raise 2.0pt\hbox{$\mathchar"13E$}}}

\def\gband{\ifmmode {g_{\rm 475}\mbox{ }}\else $g_{\rm 475}$ \fi}
\def\rband{\ifmmode {r_{\rm 625}\mbox{ }}\else $r_{\rm 625}$ \fi}
\def\iband{\ifmmode {i_{\rm 775}\mbox{ }}\else $i_{\rm 775}$ \fi}
\def\zband{\ifmmode {z_{\rm 850}\mbox{ }}\else $z_{\rm 850}$ \fi}

\begin{document}

\title{Feedback and Brightest Cluster Galaxy Formation: 
ACS Observations of the Radio Galaxy TN J1338--1942 at $\mathbf{z=4.1}$\altaffilmark{1}}

\author{Andrew W. Zirm,
R.A. Overzier,
G.K. Miley\altaffilmark{2},
J.P. Blakeslee\altaffilmark{3},
M. Clampin\altaffilmark{4},
C. De Breuck\altaffilmark{5},
R. Demarco\altaffilmark{3},
H.C. Ford\altaffilmark{3},
G.F. Hartig\altaffilmark{6},
N. Homeier\altaffilmark{3},
G.D. Illingworth\altaffilmark{7},
A.R. Martel\altaffilmark{3},
H.J.A. R\"{o}ttgering\altaffilmark{2},
B. Venemans\altaffilmark{2},
D.R. Ardila\altaffilmark{3},
F. Bartko\altaffilmark{8}, 
N. Ben\'{\i}tez\altaffilmark{9},
R.J. Bouwens\altaffilmark{6},
L.D. Bradley\altaffilmark{3},
T.J. Broadhurst\altaffilmark{10},
R.A. Brown\altaffilmark{6},
C.J. Burrows\altaffilmark{6},
E.S. Cheng\altaffilmark{11},
N.J.G. Cross\altaffilmark{12},
P.D. Feldman\altaffilmark{3},
M. Franx\altaffilmark{2},
D.A. Golimowski\altaffilmark{3},
T. Goto\altaffilmark{3},
C. Gronwall\altaffilmark{13},
B. Holden\altaffilmark{6},
L. Infante\altaffilmark{14}
R.A. Kimble\altaffilmark{4},
J.E. Krist\altaffilmark{15},
M.P. Lesser\altaffilmark{16},
S. Mei\altaffilmark{3},
F. Menanteau\altaffilmark{3},
G.R. Meurer\altaffilmark{3},
V. Motta\altaffilmark{14},
M. Postman\altaffilmark{3},
P. Rosati\altaffilmark{5}, 
M. Sirianni\altaffilmark{3}, 
W.B. Sparks\altaffilmark{3}, 
H.D. Tran\altaffilmark{17}, 
Z.I. Tsvetanov\altaffilmark{3},   
R.L. White\altaffilmark{3}
\& W. Zheng\altaffilmark{3}}


\altaffiltext{1}{Based on observations made with the NASA/ESA Hubble Space Telescope, which is operated by the Association of Universities for Research in Astronomy, Inc., under NASA contract NAS 5-26555. These observations are associated with program \# 9291}

\altaffiltext{2}{Leiden Observatory, Postbus 9513, 2300 RA Leiden,Netherlands.}

\altaffiltext{3}{Department of Physics and Astronomy, Johns Hopkins
University, 3400 North Charles Street, Baltimore, MD 21218.}

\altaffiltext{4}{NASA Goddard Space Flight Center, Code 681, Greenbelt, MD 20771.}

\altaffiltext{5}{European Southern Observatory,
Karl-Schwarzschild-Strasse 2, D-85748 Garching, Germany.}

\altaffiltext{6}{STScI, 3700 San Martin Drive, Baltimore, MD 21218.}

\altaffiltext{7}{UCO/Lick Observatory, University of California, Santa
Cruz, CA 95064.}

\altaffiltext{8}{Bartko Science \& Technology, 14520 Akron Street, 
Brighton, CO 80602.}	

\altaffiltext{9}{Inst. Astrof\'{\i}sica de Andaluc\'{\i}a (CSIC), Camino Bajo de Hu\'{e}tor, 24, Granada 18008, Spain}

\altaffiltext{10}{Racah Institute of Physics, The Hebrew University,
Jerusalem, Israel 91904.}

\altaffiltext{11}{Conceptual Analytics, LLC, 8209 Woburn Abbey Road, Glenn Dale, MD 20769}

\altaffiltext{12}{Royal Observatory Edinburgh, Blackford Hill, Edinburgh, EH9 3HJ, UK}

\altaffiltext{13}{Department of Astronomy and Astrophysics, The
Pennsylvania State University, 525 Davey Lab, University Park, PA
16802.}

\altaffiltext{14}{Departmento de Astronom\'{\i}a y Astrof\'{\i}sica,
Pontificia Universidad Cat\'{o}lica de Chile, Casilla 306, Santiago
22, Chile.}

\altaffiltext{15}{Jet Propulsion Laboratory, M/S 183-900, 4800 Oak Grove Drive, Pasadena, CA 91109}

\altaffiltext{16}{Steward Observatory, University of Arizona, Tucson, AZ 85721.}

\altaffiltext{17}{W. M. Keck Observatory, 65-1120 Mamalahoa Hwy., Kamuela, HI 96743}


\begin{abstract}

We present deep optical imaging of the $z=4.1$ radio galaxy TN J1338--1942 obtained using 
the Advanced Camera for Surveys (ACS) on-board the 
{\it Hubble Space Telescope} as well as 
ground-based near-infrared imaging data from ESO/VLT.  
The radio galaxy is known to reside within a 
large galaxy overdensity (both in physical extent 
and density contrast).  There is good evidence that 
this `protocluster' region is the 
progenitor of a present-day rich galaxy cluster.  
TN J1338 is the dominant galaxy in the 
protocluster, in terms of size and luminosity 
(in both the optical and near-infrared) and therefore 
seems destined to evolve into the brightest cluster galaxy.

The high spatial-resolution ACS images reveal several kpc-scale 
features within and around the radio galaxy.  The continuum light 
is aligned with the radio axis and is resolved into two clumps 
in the \iband and \zband bands.  
These components have luminosities $\sim 10^{9} L_{\sun}$ and sizes 
of a few kpc.  The estimated nebular continuum, scattered light, 
synchrotron and inverse-Compton scattering contributions to the 
aligned continuum light are only a few percent of the observed total, 
indicating that the observed flux is likely dominated by forming stars.  
The estimated star-formation rate for the whole 
radio galaxy is $\sim 200 M_{\odot}$ yr$^{-1}$.  
A simple model in which the jet has triggered star-formation in these 
continuum knots is consistent with the available data.

A striking, but small, linear feature is evident in the \zband aligned light 
and may be indicative of a large-scale shock associated with the 
advance of the radio jet.  
The rest of the aligned light also seems morphologically consistent with star-formation 
induced by shocks associated 
with the radio source, as seen in other high-$z$ radio galaxies (e.g., 4C~41.17).  
An unusual feature is seen in Lyman-$\alpha$ emission.  
A wedge-shaped extension emanates from the radio galaxy perpendicularly to the radio axis.  
This `wedge' naturally connects to the surrounding, asymmetric, large-scale ($\sim 100$ kpc) 
Lyman-$\alpha$ halo.  We posit that the wedge is a starburst-driven 
superwind, associated with the first major epoch of formation 
of the brightest cluster galaxy.  
The shock and wedge are examples of feedback processes due to both AGN and 
star-formation in the earliest stages of massive galaxy formation.
\end{abstract}


\section{Introduction\label{Intro}}

The most massive galaxies in the local Universe reside 
in the centers of rich clusters.  Within the context of hierarchical 
models of biased galaxy formation, the mass of a galaxy and 
its clustering properties are naturally connected via the 
initial density fluctuations \citep*[e.g.,][]{Kaiser84}.  
Therefore, not only locally, but throughout cosmic time, massive galaxies 
mark the densest regions of the Universe.  
The study of young overdensities at high redshift (`protoclusters') 
then also traces the history of the future brightest cluster galaxies.  

Many observing programs, spanning wavelengths from 
radio to X-ray, have been devoted to identifying 
galaxy overdensities over a large range of redshift 
\citep*[e.g.,][]{PLGO+96,SJEP+97,SEES+97,
RdCNG98,OPL98,RSEE+99,HNRM+99,JARON1138,PKR+00,HANC+00,DMSL+01,Francisetal01,
SEDHD02,PLOD02,DSML+02,MMQV+03,MILEYNATURE}.  
To date the most distant protoclusters have been found 
at $z \sim 5$ \citep*{Shimasakuetal03,Venemans5p2}.  
Do these very young overdensities already 
contain a dominant, massive galaxy?

There are several observational clues to the mass of a high-redshift 
galaxy.  One is the observed $K$-band magnitude, which probes the rest-frame 
optical out to $z \sim 4$.  Another is the presence of a high luminosity active nucleus 
(a supermassive accreting black-hole), 
which implies the existence of a large spheroidal 
host galaxy at least locally \citep*{MAGORRIAN,GEBHARDT,FERRARESEMERRITT}.  
High-redshift radio galaxies (HzRGs) are bright at $K$ and harbor powerful nuclei 
\citep*{Jarvisetal2001,DeBreucketal2002,Willottetal2003,JarvisMcLure2002}.  
Therefore the fields surrounding HzRGs are important targets for studying 
the earliest examples of massive galaxies and clusters.  
Using a narrow-band Lyman-$\alpha$ imaging program Miley and collaborators discovered 
an overdensity of star-forming galaxies around all four radio galaxies 
observed to sufficient depth, out to $z=5.2$ \citep*{Venemans5p2}.  
The resulting set of protoclusters is the subject of several on-going studies.  
This discovery also implies that the radio galaxies are the seeds 
of brightest cluster galaxies.

Brightest cluster galaxies (BCGs) are the most massive galaxies 
known in the local Universe; having stellar masses in excess 
of $10^{12}M_{\odot}$ \citep*{JFK96,Bernardietal2003}.  
The luminosities and sizes 
of brightest cluster galaxies are not drawn from 
the same distributions as the majority of the galaxy 
population \citep*{OegerleHoessel91}.  BCGs at low redshift 
lie on the extrapolated Fundamental Plane of elliptical galaxies \citep{OegerleHoessel91}.  
The surface-brightness profiles of some BCGs, the cD galaxies, extend out to 
hundreds of kiloparsecs.  These shallow power-law stellar envelopes blur the 
distinction between the galaxy and the diffuse intracluster light.  
Such extreme sources are clearly very useful laboratories for 
studying the processes inherent to massive galaxy and 
cluster formation.  In fact, several authors have shown that 
the observed build-up of BCGs can provide key constraints on the 
hierarchical theory of galaxy formation \citep*{A-SBK98,BurkeCollinsMann00}.  
The present discrepancies between the predicted (using semi-analytic 
models) and observed abundance of massive galaxies imply that 
fundamental processes are not being accounted 
for in the current models \citep*{Coleetal2000,Baughetal2003,Somervilleetal2004}.  
One possibility to solve this discrepancy is to postulate the existence of 
strong interactions between accreting black-holes, 
star-formation and their host galaxy and surroundings (i.e., ``feedback'').  

Galaxies which host powerful radio sources are 
peculiar in several respects.  The most striking property 
is the radio-optical `alignment effect'.  
This effect is the strong tendency of the 
rest-frame ultraviolet continuum light of the radio host 
to be aligned with the axis defined by the 
radio source.  Several explanations 
for this behavior have been put forward: recent star-formation 
induced by the radio jet, nebular continuum from emission line 
gas which is photo or shock ionized by the AGN, light from the central engine 
scattered into the line of sight by either dust or electrons, and  
inverse-Compton scattering of the microwave background or other local photon 
fields \citep*{McCvBSD87,ChambersMileyvanBreugel87,Daly1992,Dicksonetal95}.  
These explanations 
of the alignment, but particularly the jet-induced star-formation, 
are excellent examples of feedback on the galaxy formation process.

The giant ($\sim$100 kpc) Lyman-$\alpha$ emitting 
halos surrounding distant radio galaxies 
may be an observable consequence 
of feedback from galaxy formation \citep*[e.g.,][]{vanOjiketal1997,ReulandHalos03}.  
The enormous line luminosities of these objects, often 
in excess of $10^{44}$ ergs s$^{-1}$, imply they are massive 
reservoirs of gas \citep*{JARON1138,BLOBS}.  
What is the origin of this gas; outflow from the galaxy or  
infall of primordial material?
There are several pieces of circumstantial 
evidence that these halos are connected with the 
AGN.  The halos are often aligned with the FRII 
radio axis, and in some cases Ly$\alpha$ 
emission is directly associated with radio structures \citep*{JARON1138_2}.  
Perhaps photoionization by the AGN or shocks due to the 
radio source expansion are responsible for the extended 
line emission.  
Spectroscopically, Ly$\alpha$ absorption is seen in addition to the 
bright halo emission \citep*[e.g.,][]{vanOjiketal1997,Wilmanetal2004abs}.  
There is a good correlation between the amount of absorption 
and the size of the radio source \citep*{Jarvisetal2001abs}.  
A possible scenario is that a neutral hydrogen shell initially 
surrounds the radio source, but is subsequently ionized during the 
growth of the radio source.  There are also some halos known which 
do not contain a bright radio source \citep*{BLOBS} suggesting that perhaps 
the halo phenomenon is associated with the more general processes of 
galaxy formation, rather than being specific to active nuclei.

The radio galaxy TN J1338--1942 \citep*[$z=4.1$;][]{1338SPEC99} 
resides in one of the youngest protoclusters known \citep*{VKM+02,MILEYNATURE}.  
This galaxy lies within a large Ly$\alpha$ halo which shows unusually 
asymmetric morphology when compared to other similar radio sources \citep{VKM+02}.  
In this paper we present high spatial-resolution {\it HST} Advanced Camera for Surveys 
(ACS) imaging of this radio galaxy.  
These data provide the clearest view of a young brightest cluster 
galaxy to date.  Images in four broadband filters ($\gband$,$\rband$,$\iband$ and $\zband$) have 
been obtained.  The resulting magnitudes and colors have been used 
to apply the `Lyman-break' technique to select galaxies 
at the same redshift as the radio galaxy \citep{MILEYNATURE}.   
The exquisite spatial resolution of {\it HST}/ACS allows us to study the 
detailed morphology of the radio galaxy.  Using these data we present a 
scenario which describes the observed morphology (both continuum and Ly$\alpha$) 
and the measured kiloparsec-scale colors and magnitudes within a self-consistent 
formation framework for TN J1338--1942 (hereafter TN J1338).

This paper is structured as follows: we describe the observations 
and data reduction in \S~\ref{Obs}, we present the results of our 
analysis of the combined multi-wavelength dataset in \S~\ref{Results}, and discuss 
these results in \S~\ref{Discussion}.  
We adopt the `concordance' cosmology \citep*{SpergelWMAP} with 
$\Omega_{m}=0.27$, $\Omega_{\Lambda}=0.73$ and $H_{0} = 71$ km s$^{-1}$ Mpc$^{-1}$.  
Within this cosmology the angular scale at the redshift 
of TN J1338, $z=4.1$, is 7.0 kpc arcsec$^{-1}$.  We use the AB magnitude 
system \citep*{AB} except where noted.

\section{Observations and Data Reduction\label{Obs}}

\subsection{ACS Imaging}

The ACS data of TN J1338 were taken with two primary 
goals in mind: first, to enable color-selection of faint 
protocluster members using the `Lyman-break' technique and 
second, to investigate the detailed morphological structure 
of the brightest protocluster galaxies, including the radio galaxy.
To achieve these goals, images were taken in 4 broad-band filters: 
\gband\ (F475W), \rband\ (F625W), 
\iband\ (F775W) and \zband\ (F805LP).  
The \gband-band is below the Lyman-break for 
galaxies at $z \simgt 4.1$.  The \rband-band contains 
(and for the case of the radio galaxy, is dominated by) the 
Lyman-$\alpha$ emission line.  Both the \iband and \zband bands 
are relatively unaffected by bright emission lines 
(see Table~\ref{tab:apmag}).  However, we note that one emission line, 
CIII]$\lambda1909$, may be affecting the \zband-band 
morphology of the radio galaxy.  
The \gband, \rband and \iband observations were 
carried out between 8 and 12 July 2002 and the \zband images between 
11 and 12 July 2003 with the Wide Field
Channel of the ACS. The total observing time of 18 orbits was split between 
the broad-band filters 9400 seconds in \gband, 9400 s in \rband, 
11700 s in \iband, and 11800 sec in \zband.  
Each orbit was split into two 1200 sec exposures to facilitate the removal of cosmic rays.  
$\gband$ dropouts 
were selected using the \gband, \rband and \iband bands as described 
in \citet{MILEYNATURE} and Overzier et al., in prep.  This effort was very successful 
and confirmed the presence of a galaxy overdensity around TN J1338.  
In this paper we 
present the first discussion of the radio galaxy itself and its role 
as the dominant protocluster galaxy.

The ACS data were reduced using the ACS pipeline science investigation software
\citep*[{\it Apsis};][]{APSIS}, developed for the ACS Guaranteed Time
Observation (GTO) program. After the initial flat-fielding of the images
through CALACS at STScI, the {\it Apsis} processing steps include the
empirical determination of image offsets and rotations, the rejection
of cosmic rays, the combining of images through {\it drizzling}, and object
detection and photometry using SExtractor \citep{bertin96}. The final
images have a scale of $0\farcs05$ per pixel, and ($2\sigma$) limiting
(AB) magnitudes of 28.46 (\gband), 28.23 (\rband), 28.07 (\iband), and 27.73 
(\zband) in 0.2-arcsec$^2$ apertures (corrected for Galactic extinction).  

\subsection{VLT Optical Imaging and Spectroscopy}

Deep VLT/FORS2 images ($6\farcm8 \times 6\farcm8$) of the TN J1338 field were made in March 2001 
in both the broad-band $R$ filter ($\lambda_{\rm c} = 6550$ \AA, 
$\Delta \lambda_{\rm FWHM} = 1650$ \AA) and a custom 
narrow-band filter ($\lambda_{\rm c} = 6195$ \AA, 
$\Delta \lambda_{\rm FWHM} = 60$ \AA) to target the redshifted Lyman-$\alpha$ emission 
line \citep*{VKM+02}.  The $1\sigma$ limiting surface-brightnesses are 28.6 and 29.2 
per arcsec$^2$ for the narrow and broad bands respectively.  
These images were used to identify candidate Ly$\alpha$ emitting galaxies 
\citep*{VKM+02}.  For the current paper these images are used to 
elucidate the larger scale structure of the Lyman-$\alpha$ emission and to compare it 
to the structures seen in the ACS images.

A follow-up spectroscopy program using FORS2 in multiobject mode was carried out 
in May 2001.  These spectra have a dispersion of 1.32 \AA\ pixel$^{-1}$ (using 
the 600RI grism) and cover the wavelength range from 5300 to 8000 \AA.  
Candidate Ly$\alpha$ galaxies were placed on two slit-masks 
\citep{VKM+02}.  The radio galaxy itself was included on both slitmasks providing a 
deep spectrum along the radio axis which covers both the Ly$\alpha$ and CIV$\lambda1549$ 
emission lines.  

\subsection{VLT NIR Imaging}

$K_S$-band images of TN J1338 were obtained in two separate observing 
runs.  One on March 24--26 2002 collected 2.1 hours of total exposure time 
using ISAAC on UT1 of the Very Large Telescope (VLT). 
The second run, using the same instrument, was done in service mode at VLT 
between the nights of 27 May and 13 June 2004.  The total exposure time for this run 
was 5.7 hours.
All the data were processed, sky-subtracted and combined using 
the XDIMSUM package within IRAF\footnote{IRAF is distributed by the National Optical Astronomy Observatories,
which are operated by the Association of Universities for Research
in Astronomy, Inc., under cooperative agreement with the National
Science Foundation.} \citep*{IRAF}.  
The final image has a scale of 0.148\arcsec per pixel, a seeing of $0\farcs5$,
and a ($2\sigma$, 1 sq. arcsec aperture) limiting magnitude of 25.6.  
ISAAC has some  geometrical distortion across the face of the 
detector.  We have not corrected the distortion in detail because 
the $K_S$ morphology of the radio galaxy matches features seen in 
the continuum ACS observations.

\begin{figure}
\epsscale{0.6}
\plotone{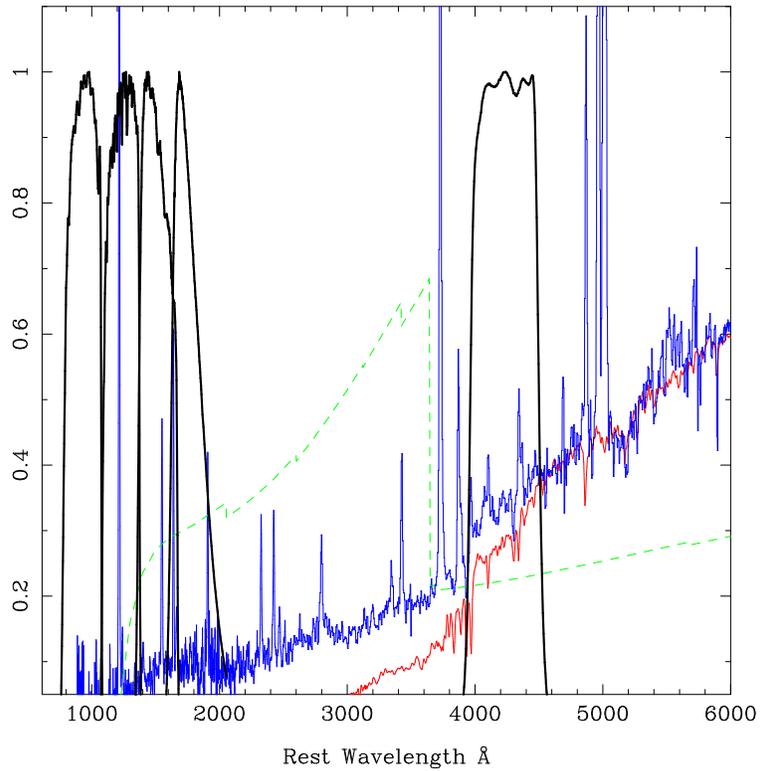}
\caption{Rest-frame wavelengths covered by the ACS and ground-based near-infrared filters.  
Shown left to right are the filter curves for the F475W, F625W, F775W, F850LP and $K_S$ 
bandpasses.  Also shown for comparison are the characteristic spectra of the nebular continuum emission 
(green), high-redshift radio galaxies in general \citep[blue;][]{McCarthy93} and an older (1 Gyr) stellar 
population (red).  
\label{fig:seds}}
\end{figure}

The $K_S$-band is the only band which probes wavelengths longward of the 4000 \AA\ break, 
beyond which an old stellar population should dominate the emergent flux 
(see Figure~\ref{fig:seds}).  It should be noted, however, that the bandpass is not entirely 
at $\lambda > 4000$ \AA.  However, these data still provide a crucial point 
on the spectral energy distribution (SED) of the radio galaxy.

\subsection{Radio Imaging}

\begin{figure}
\epsscale{1.0}
\plotone{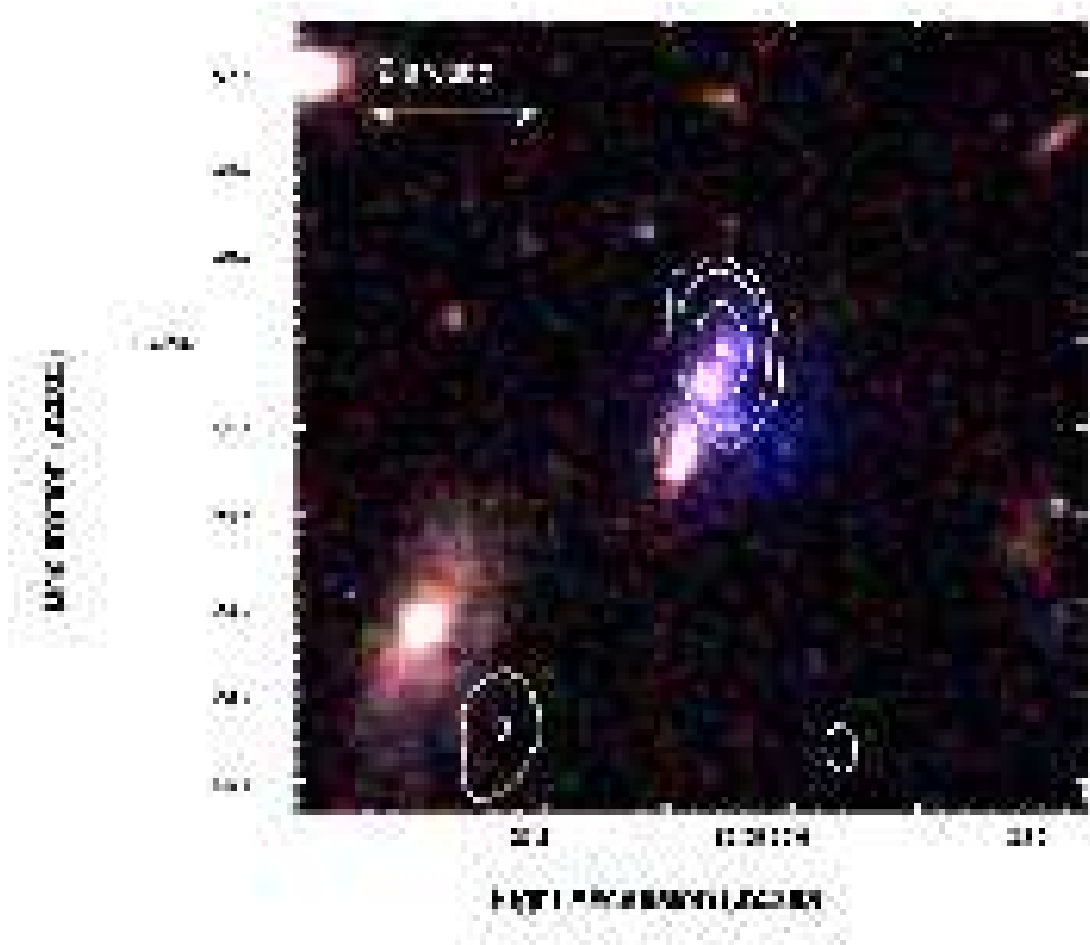}
\caption{Color composite image of the \rband, \iband and \zband images from 
ACS, with the VLA 5 GHz radio map overlaid.  Notice the blue `wedge' emanating from the 
south-western side of the radio galaxy, this appears only in the \rband\-band and 
is likely due entirely to Lyman-$\alpha$ emission.  The rest of the emission is 
clearly aligned with the axis defined by the two radio lobes.  The continuum in the 
\iband and \zband bands is dominated by the two clumps along the same 
axis.  
\label{fig:colorradio}}
\end{figure}

The radio source TN J1338 was originally selected because of its ultra-steep spectrum 
(between 365 MHz and 1.4 GHz) which has been shown to be an indicator of high 
redshift \citep*{DBvBRM00}.  
The first radio data were culled from the Texas and NVSS catalogs \citep*{TEXAS,NVSS}.  
Follow-up observations were made with the VLA at 4.71 and 8.46 GHz in March 1998 
\citep*{Pentericcietal00}.  The noise levels are 25 and 50 $\mu$Jy beam$^{-1}$ for 
the 8 and 5 GHz maps, respectively.  The resolution is $0\farcs23$ for the 8 GHz 
and $0\farcs43$ for the 5 GHz map.  
These are the primary radio data used in this paper (see overlay in Figure~\ref{fig:colorradio}).  
The radio source has 
three distinct components at both these frequencies; the 
northwest ($S^{\rm NW}_{4.7 {\rm GHz}} = 21.9$ mJy) and southeast 
($S^{\rm SE}_{4.7 {\rm GHz}} = 1.1$ mJy) lobes (separated by $5\farcs5$) and 
the likely radio core ($S^{\rm core}_{4.7 {\rm GHz}} = 0.3$ mJy) 
located very close ($1\farcs4$) to the northwest lobe \citep*{1338SPEC99}.  
The radio source is highly asymmetric, with the NW lobe nearly 20 times brighter 
at 4.7 GHz than the SE lobe.  The radio asymmetry may indicate an asymmetry 
in the ambient medium \citep*{McCvBK91}.  

\subsection{Image Registration}

The center position, orientation and angular-resolution all differ between the 
ground-based and space-based data sets.  To facilitate comparison amongst these 
datasets we registered them all to a common reference frame and pixel scale.  
The ACS images were registered with respect 
to each other in the ACS GTO pipeline.  
We chose the ACS \rband frame as this common grid.  
For the ground-based $R$ and $K_S$ data this 
registration requires interpolation from plate-scales of $0\farcs201$ and $0\farcs148$ 
to $0\farcs05$ per pixel respectively.  The shifts, rotations and 
rebinning were all done in a single interpolation step using the 
IRAF tasks {\it geomap} and {\it geotran}.  We fitted a general coordinate transformation 
using 2nd order polynomials in both axes.  The rms deviations of the data from the fits 
were of order 0.3 input pixels ($0\farcs04$) for the $K_S$ image and 
0.07 input pixels ($0\farcs01$) for the 
FORS2 narrow and broad band images.  The calculated transformations were done using 
`sinc' interpolation within {\it geotran}.  
At least 15-20 unsaturated stars were matched within the ACS \rband-\-band 
and each of the ground-based images to calculate the appropriate 
transformations.  We applied these transformations to the ground-based $R$, narrow-band 
and $K_S$ images.  

The optical/near-infrared frame  is defined by stellar positions and that of the radio image 
is defined by the positions of radio point sources (quasars).  
The radio image of the TN J1338 field is sparse, so a direct matching of 
sources will not produce a robust transformation between the two frames; particularly 
when we do not want to use the radio galaxy itself.  
Therefore the accuracy limit to the radio-optical registration is determined solely 
by the systematic error in the optical reference frame.  To better quantify 
this we have used two different optical frames, the USNO and the GSC-2.0, and 
compared the astrometric solutions using both to the radio data.  This gives us a conservative 
amplitude of $0\farcs3$ to the error on the position of the radio core with respect to the 
optical structures in the ACS and ground-based data.  
Therefore we can confidently associate the radio core with the region near the peak of the 
$K_S$-band flux and at one tip of the ACS galaxy. 

\subsection{Continuum subtraction}

To model the continuum in the \rband we have used a power-law extrapolation from 
the relatively emission-line free \iband and 
\zband bands.  While these filters are not strictly line-free due to the presence of 
CIV$\lambda1549$ in the \iband\-band, HeII$\lambda1640$ in both passbands 
and, to a lesser extent, CIII]$\lambda1909$ in 
the \zband\-band, none of these lines are expected to 
dominate the continuum (based on spectroscopic data, see Table~\ref{tab:apmag}).  
We assume that the continuum follows a simple power-law in 
$F_{\lambda} (\propto \lambda^{\beta})$ that extrapolates 
through to the ACS \rband bandpass, VLT $R$ and narrow-bands.  
We have accounted  for the intergalactic 
absorption shortward of the emission line and the relative throughputs of the filter curves.  
Light at wavelengths shorter than Lyman-$\alpha$ is easily absorbed by neutral 
hydrogen located between the source and the observer.  This will greatly affect the 
amount of continuum light detected in any bandpass shortward of the emission line.  
We have therefore adopted the model of the intergalactic hydrogen optical 
depth presented in \citet*{Madau95}; 
$\tau_{\rm eff} = 0.0036 (\frac{\lambda_{\rm obs}}{\lambda_{\beta}})^{3.46}$.
This is the optical depth due to Lyman-$\alpha$ forest lines, and not due to higher 
order Lyman series or metal lines.  
These two opacity sources make negligible contributions to the total optical 
depth at this redshift \citep*{Madau95}.  
We integrate this attenuation over the filter curve shortward of Ly$\alpha$ to determine 
the amount of continuum flux absorbed in the intergalactic medium (IGM).  This correction 
decreases the amount of continuum by 23.9\%, 16.9\% and 36.2\% 
in the ACS \rband, VLT $R$ and VLT narrow-bands respectively.
In addition, the continuum flux detected in the images is affected by the filter 
throughput curves.  The extrapolation from continuum measured in some filters 
(in our case, the ACS \iband and \zband) to other bandpasses depends 
on the relative filter curves.  We have used the total integrated 
throughputs to correct for the differential sensitivities.  
The final value of $\beta$ for the entire radio galaxy is 
$-1.32$ (or $\alpha = 0.62$, where $F_{\nu} \propto \nu^{\alpha}$).  

\section{Results\label{Results}}

\begin{figure}
\plotone{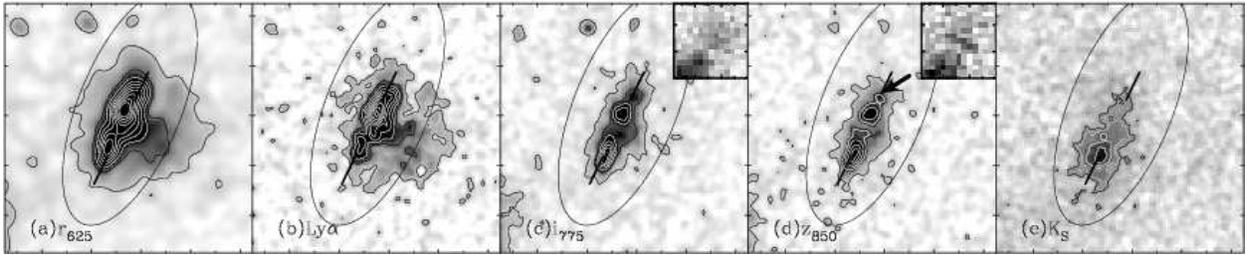}
\caption{Four optical and one near-infrared images used in this paper.  
North is up and East to the left.  Each cutout is $4\farcs5$ on a side.  
From left to right: the \rband ACS image, 
the same image but continuum-subtracted, the \iband 
ACS image, the \zband ACS image and finally the $K_S$ image from VLT.  
All the ACS images have been smoothed with a 1.5 pixel Gaussian kernel.   
The arrow in panel (d) marks the linear feature which is likely a 
large-scale shock, which is also shown unsmoothed in the inset.  The two lines 
indicate the radio axis and the ellipse is the Kron aperture used for 
photometry of the entire radio galaxy.
\label{fig:multi}}
\end{figure}

The ACS and VLT images are shown in Figure~\ref{fig:multi}.  
The rest-frame ultraviolet morphology of TN J1338
is complex and multi-faceted.  The radio galaxy exhibits the usual 
alignment between its continuum, line emission 
and radio axis \citep*{ChambersMileyvanBreugel87,McCvBSD87,BLR98}.  
Two kiloparsec-sized clumps along the radio axis dominate the continuum structure 
in the \iband and \zband bands.  This is similar to many 
powerful 3CR radio galaxies at $z \sim 1$ \citep*{BLR98,THESIS,NICPAP}.  
In the \rband\-band, the ACS image reveals both concentrated and diffuse 
Lyman-$\alpha$ emitting regions.  
In this section we quantify the line and continuum flux distributions 
for the radio galaxy by performing photometry in a set of varied apertures.  

\subsection{Estimating the Contributions to the Aligned Light\label{Ana:Align}}

The presence of a powerful radio source has several effects on 
the emitted spectrum.  As mentioned above, at least three of these 
effects tend to align the observed continuum emission with the axis defined by 
the double radio lobes.  
Any remaining continuum flux we attribute to a young stellar population.  

Ionized gas not only emits emission lines, but also continuum 
photons from two-photon recombination, bremsstrahlung and standard 
recombination.  For a given gas temperature and density, 
this nebular continuum spectrum can be calculated (see Figure~\ref{fig:seds}).  
Observationally, the normalization of the nebular spectrum is determined 
by using the emission line spectrum.  
For TN J1338--1942 we have a spectrum which was taken along the radio axis 
and spatially-averages the entire optical extent of the radio galaxy (RG).  
Ideally one would measure Balmer recombination emission lines.  
These lines directly correspond to the nebular continuum.  
At $z=4.1$, these lines have shifted out of the optical 
window visible from the ground.  
As a `Balmer proxy' we use the HeII$\lambda1640$ emission line \citep*{Vernetetal2001}, 
which has a flux of $\approx 1.7 \times 10^{-16}$ ergs s$^{-1}$ cm$^{-2}$ 
(2004, C. De Breuck, private communication).  
Assuming the HeII/H$\beta$ ratio ($=3.18$) from a high-redshift radio galaxy 
composite spectrum \citep*{McCarthy93}, we estimate the H$\beta$ flux to 
be $5.4 \times 10^{-17}$ ergs s$^{-1}$ cm$^{-2}$.  If these emission lines arise in a 15,000 K 
gas, at low enough densities where collisional de-excitation is negligible, 
then the corresponding nebular continuum flux densities in the \rband, \iband 
and \zband bands are $1.5 \times 10^{-31}$, $3.1 \times 10^{-31}$ and 
$3.8 \times 10^{-31}$ ergs s$^{-1}$ cm$^{-2}$ Hz$^{-1}$ respectively.  
So, at its brightest the nebular continuum averaged 
over the entire galaxy is only \zband(AB) =27.4, 
much fainter than even the individual components of TN J1338.  
We subtract the nebular continuum from all quoted magnitudes and fluxes by 
scaling the subtracted amounts for the sub-components by their estimated Ly$\alpha$ flux.

The same population of electrons that is responsible for the radio 
synchrotron emission can also inverse-Compton (IC) scatter ambient 
photon fields.  The primary seed photons are those making up the 
cosmic microwave background (CMB), with secondary contributions from 
the synchrotron photons themselves (synchrotron self-Compton; SSC) 
and other AGN emission.  By assuming equipartition 
to calculate the magnetic field in the northern radio lobe of TN J1338, we 
find a $B$-field on the order of a few hundred micro-Gauss.  
Using the extrapolation to the rest-UV as calculated by \citet*{Daly1992}, 
this translates to an IC-CMB contribution of only a 
few $\times 10^{-31}$ ergs s$^{-1}$ cm$^{-2}$ Hz$^{-1}$, 
or fainter than 28 magnitudes (AB).  The estimated energy density in the 
synchrotron field is similar to that of the CMB at this high-redshift, so SSC 
adds a similar amount to the aligned continuum.  These two contributions combined 
are negligible.  As with the spectroscopy above, 
these values are spatially-averaged over the entire galaxy.  

Optical spectropolarimetry of the entire galaxy reveals a 
polarization of $ 5 \pm 3\%$ (2004, C. De Breuck, private communication).  
This value is much lower (about a factor of two) than similar measurements 
of $z \sim 1$ RGs 
\citep*[e.g.,][]{Deyetal1996,Cimattietal1996,Cimattietal1997,Solorzanoetal2004}, 
but similar to other powerful radio galaxies 
at $z \sim 4$ \citep*{Dey41p17}.  The amount of scattered light is related 
to the percent polarization by the intrinsic polarization, $P_i$.   $P_i$ depends on 
the type of scatterer (either dust or electrons) and the geometry of the 
scattering, neither of which is well-constrained for TN J1338.  Therefore we 
can only put the lower limit on the percentage of scattered light of 5\%.  However, 
we also note that for $z \sim 1$ 3CR radio galaxies the observed polarizations are 
high and suggest that $P_i$ is also high.  
We therefore conclude that for TN J1338 the aligned 
continuum contains some scattered light ($\sim 10\%$), but is substantially 
diluted by unpolarized sources.  
We have already shown that neither the nebular continuum nor the IC 
scattering can account for this dilution.  

Clear evidence for the presence of young stellar populations in radio galaxies 
has proven difficult to find.  The same spectral region, the ultraviolet, 
where massive stars are brightest coincides with the 
bright region of the AGN spectral energy distribution.  Therefore, the detection depends on 
being able to find stellar-specific features in very deep spectra 
\citep*[e.g., the SV$\lambda1502$ photospheric absorption 
feature in 4C~41.17 at $z=3.8$][]{Dey41p17}.  
Existing spectra of TN J1338 do not show such features.  In this case 
the presence of young massive stars must be inferred from the UV excess 
after subtracting the other known contibutors to the aligned light.  
As we have shown above, there seems to be such an excess in TN J1338, 
which implies the existence of many young stars.  We examine this result 
further below.

\subsection{Large Aperture Photometry\label{Ana:Aper}}

To perform integrated photometry over the whole galaxy we used 
a ``Kron'' aperture \citep*{Kron80} 
to approximate a total galaxy magnitude in the ACS and VLT images.  
This enables easy comparisons between the space and ground-based observations of TN J1338.  
This aperture is optimized to be as large as possible while still retaining a 
high signal-to-noise ratio.  The magnitudes were determined using 
SExtractor \citep*{bertin96}.  
The {\it Apsis} pipeline ``deblends'' the radio galaxy into 
two distinct objects.  We therefore reset the SExtractor deblending parameters 
to maintain the radio galaxy as a single object.  
We overplot the aperture in Figure~\ref{fig:multi}.  The `total' magnitudes within 
this aperture are listed in Table~\ref{tab:apmag}.  

The radio galaxy is $>1.4$ magnitudes brighter 
than the next brightest ``dropout'' or 
Lyman-$\alpha$ emitting galaxy (within similarly defined apertures).  
This is true even in the $K_S$ band where the light from older stars is presumed to 
dominate the emission rather than processes related to the AGN.  Therefore the radio galaxy 
is securely identified as the dominant, and likely most massive, galaxy within 
the protocluster and the probable progenitor of a present-day brightest cluster galaxy.  

\subsection{Rest-frame Optical Surface-brightness Profile\label{Ana:Optical}}

\begin{figure}
\plotone{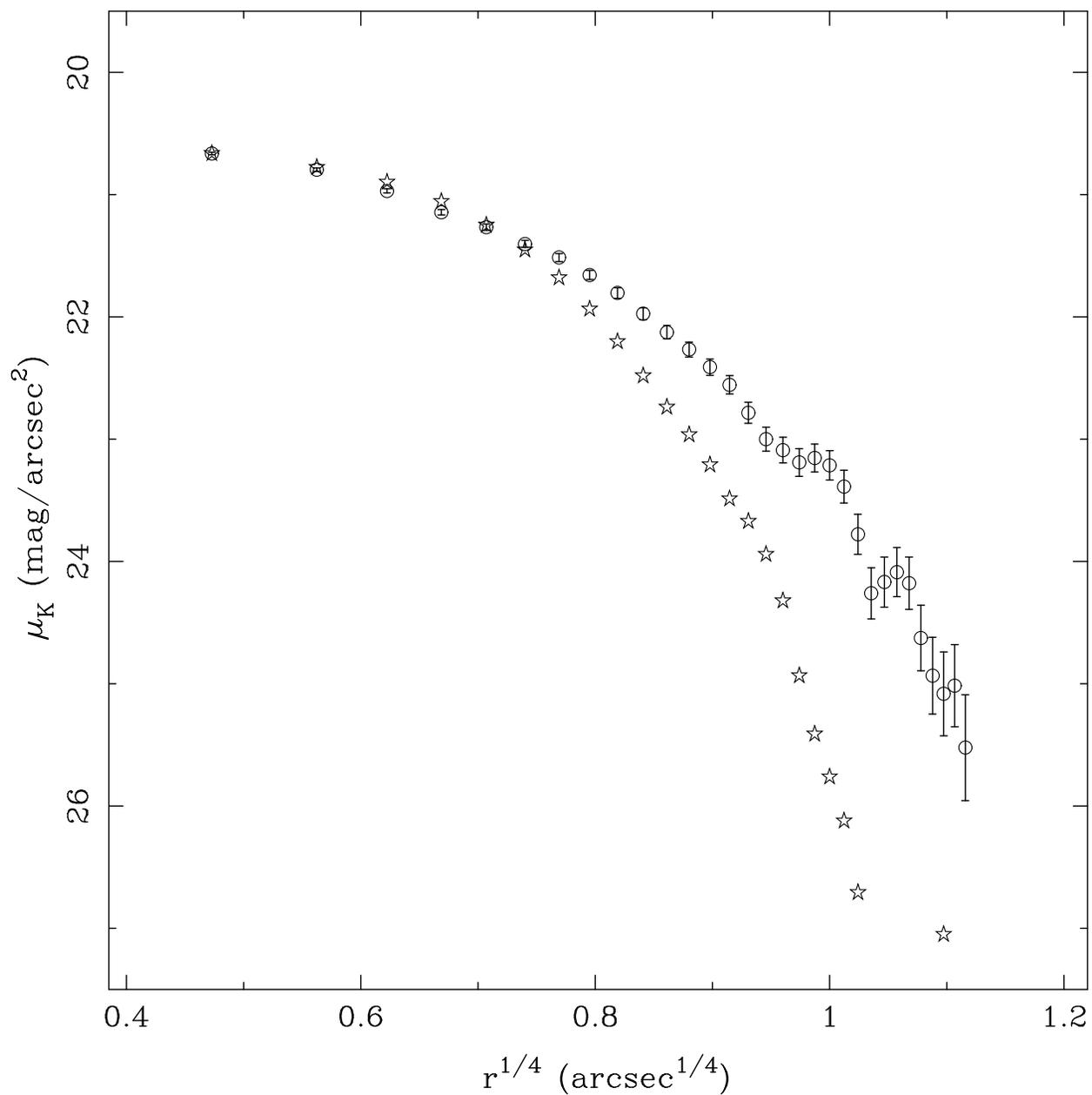}
\caption{$K_S$--band surface-brightness profile for TN J1338 extracted in 
circular apertures (open circles) along with that of a star measured in the same manner (stars).  
The radio galaxy is clearly extended and there is some indication that the profile 
may follow a de Vaucouleurs law.  However, these data does not have sufficient spatial 
resolution to decisively determine the form of the surface-brightness profile.
\label{fig:NIR}}
\end{figure}

The $K_S$-band samples the rest-frame optical continuum 
emission, mostly longward of the 
4000 \AA\ break, and shortward of the potentially 
bright [OIII]$\lambda\lambda4959,5007$ and 
H$\beta$ lines.  At these wavelengths it is likely that 
the galaxy luminosity is dominated 
by older, low-mass stars.  We show the final ISAAC image of the radio galaxy 
in panel (e) of Figure~\ref{fig:multi} and 
the raw surface-brightness profile of the radio galaxy and a star in this band 
in Figure~\ref{fig:NIR}.  This profile is centered on the peak of the $K_S$-band 
light.  It is clear that the galaxy is resolved, but 
it is not possible to measure the size of the galaxy robustly from these 
data.  A simple $r^{\frac{1}{4}}$-law fit roughly estimates the 
effective radius at $0\farcs7$ or $\sim 5$ kpc.  This is considerably smaller than 
the corresponding radii of local BCGs \citep[e.g.,][]{GLCP96}.  
In fact, the observed morphology in the $K_S$ image is very similar to that seen 
in the ACS \iband-band.  
The galaxy is highly elongated and is nearly perfectly aligned with the 
radio axis.  There are no bright emission lines within the bandpass, and none 
of the continuum processes discussed above are bright in $K_S$.  
Therefore, we conclude that the stellar distribution is aligned 
with the radio axis.  

\subsection{Multi-component Decomposition and Diffuse Light\label{Ana:Stukjes}}

\begin{figure}
\plotone{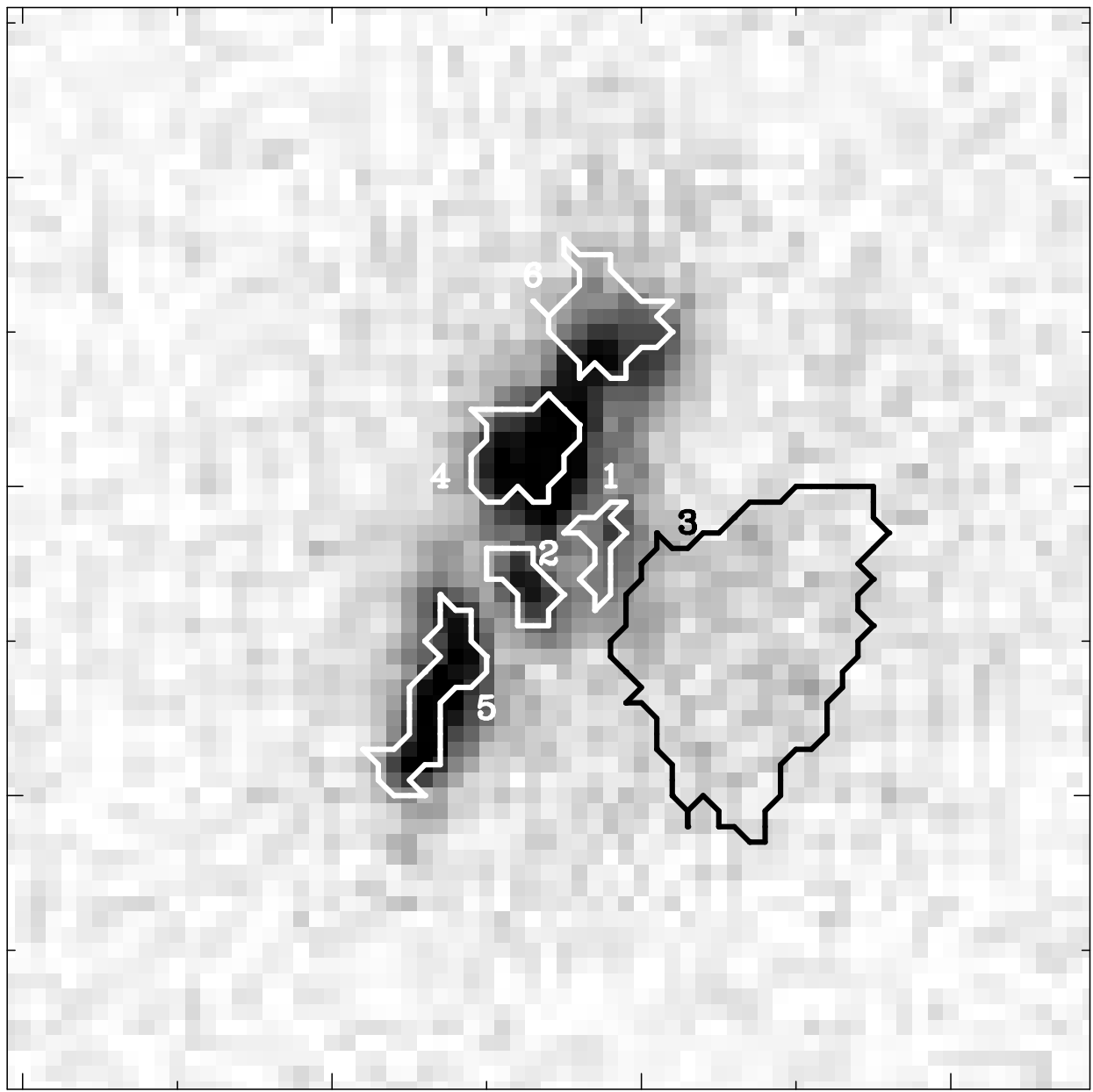}
\caption{Segmentation map of the galaxy showing the labelled individual clumps.  North is up and 
East to the left.  This cutout is $3\farcs5$ on a side, with major tickmarks every arcsecond.  
\label{fig:SegMap}}
\end{figure}

The rest-frame ultraviolet morphology of 
TN J1338 can be decomposed into several 
discrete components.  To better understand the morphology of 
the galaxy we have extracted multi-band photometry of these individual components.  
We divided the radio galaxy into distinct regions 
both in the line-dominated \rband-band and in the 
continuum-dominated \iband-band.  The resulting segmentation map is 
shown in Figure~\ref{fig:SegMap}.  The regions are numbered 1 thru 6.  
Regions 1, 2 and 3 (`wedge') are found only in 
the emission-line dominated \rband-band.  Conversely, regions 4 and 
and 5 are most prominent in the continuum bands (\iband and 
\zband).  Finally, region 6 contains a linear feature 
in \zband.

For each region we have measured the magnitude in each ACS band and 
in the ACS Ly$\alpha$ (continuum subtracted \rband) image.  Using these values we have derived the 
UV continuum slope and line flux independently for each portion of the galaxy.
In addition, we have measured the magnitudes in the $K_S$-band for the two 
primary continuum clumps visible in the ACS data (regions 4 and 5).  
The $(\iband - K_S)_{\rm AB}$ color for region 5 is $\sim 0.2$ magnitudes redder 
than region 4.  
The sum of the light contained within these clumps is about one magnitude fainter in the 
\iband-band than the total radio galaxy, indicating that there is some 
``diffuse'' light detected in these high angular-resolution images.  
This remains the case in the \zband-band with the extended light 
apparently having the same color as the mean color of the entire galaxy.
The magnitudes and Ly$\alpha$ fluxes are presented in Columns 3-6 of Table~\ref{stukjetab}.

\subsection{Azimuthal Binning}

\begin{figure}
\plotone{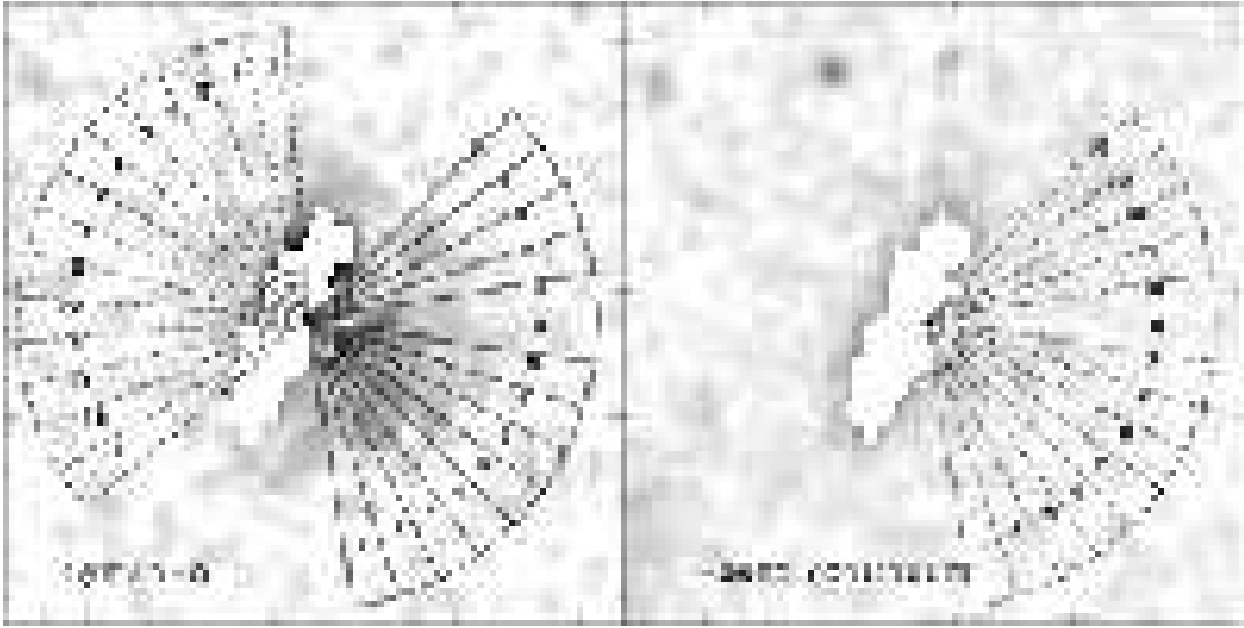}
\caption{Azimuthal bins used to measure the profile shown in Figure~\ref{fig:WedgeHist} 
overplotted on the continuum-subtracted \rband (left) and \iband (right) images.
\label{fig:Wedges}}
\end{figure}

\begin{figure}
\plotone{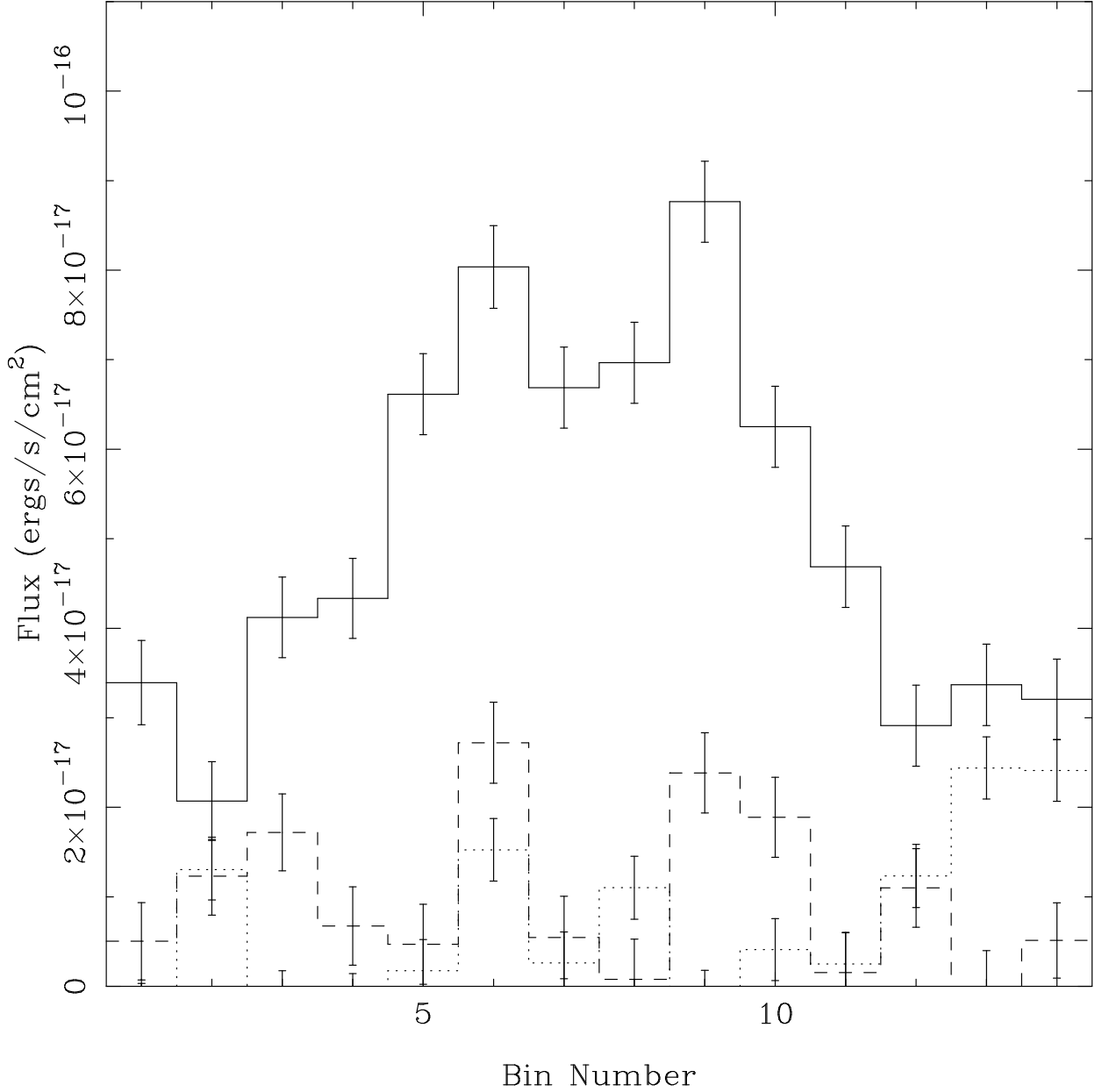}
\caption{Azimuthal profile of the wedge (solid line), `anti-wedge' (dashed line) and 
\iband continuum (dotted line).  Note the sharp cutoff on either side of the wedge and the 
possible sub-structure (radial `filaments') in the profile. \label{fig:WedgeHist}}
\end{figure}

While the aligned light is a common feature 
of high-redshift radio galaxies, the wedge of extended 
line-emission to the southwest of the radio galaxy 
(region 3 in Figure~\ref{fig:SegMap}) is unusual.  
We have measured the azimuthal surface-brightness distribution 
of this feature by extracting photometry in angular bins.  
The bins are shown in Figure~\ref{fig:Wedges}.  
The angular regions are sized and placed to cover the entire visible extent 
of the wedge and to be small enough to accentuate the 
internal structure of the feature.  The flux in each bin was summed 
using the IRAF task `polyphot'.  The corresponding errors were calculated 
by performing the same aperture photometry on the error maps.  
The pixels within the radio galaxy 
(as defined by the extent of the \iband-band continuum) were masked out.  
The resulting flux histogram is shown in Figure~\ref{fig:WedgeHist} for 
the wedge (solid line), the opposite side of 
the galaxy in the line image (`anti-wedge', dashed line) 
and for the \iband-band continuum in the wedge region (dotted line).  

The bins covering the wedge show a clear excess of flux with respect to 
both the opposite side of the galaxy and to the continuum.  
There also appears to be some structure to the azimuthal profile.  
The azimuthal profile has relatively sharp-cutoffs at either 
side.  There is no evidence for limb-brightening toward the edges of the wedge.  
However, the spatial resolution ($\sim 1.5$ kpc/bin) is insufficient to 
resolve a thin shell of limb-brightened emission.  

\subsection{Semi-circular Annuli}

\begin{figure}
\plotone{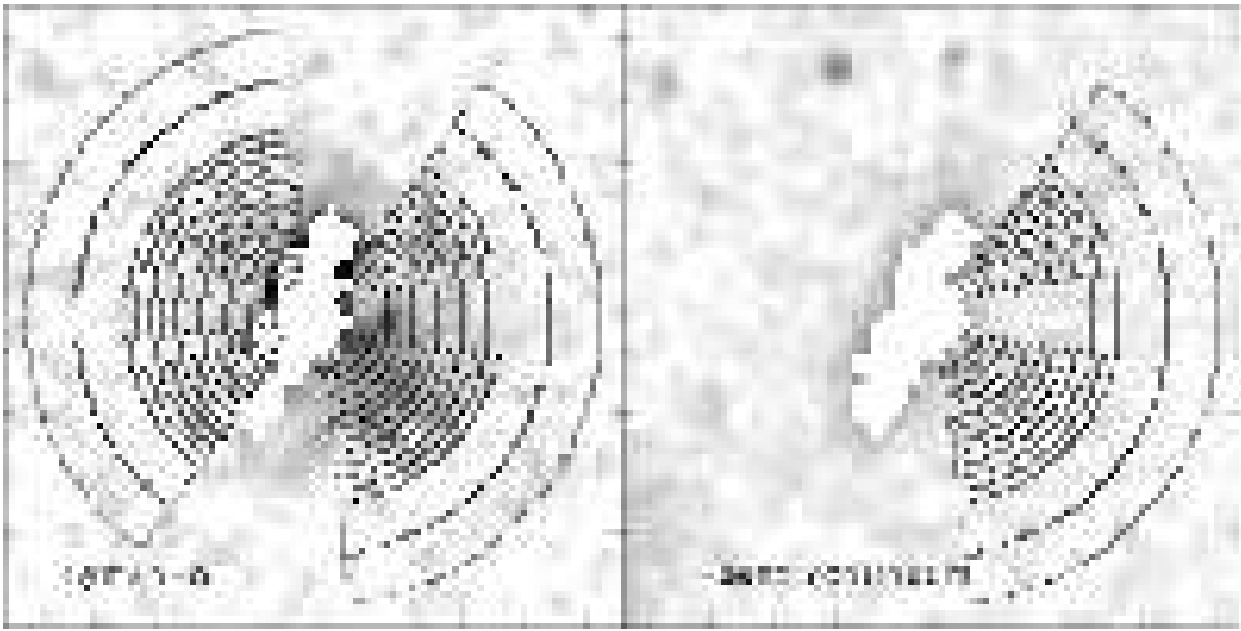}
\caption{Semi-circular annuli used to measure the radial profile of the wedge in Ly$\alpha$ (left).  
The same annuli were also used on the `anti-wedge' side of the galaxy, and on the 
$i$-band continuum image (right).  North is up and East to the left.  Each cutout is 
$5\arcsec$ on a side, with major tickmarks separated by 1 arcsec.  
The radial dependence of all three are shown in 
Figure~\ref{fig:profile}.\label{fig:annuli}}
\end{figure}

\begin{figure}
\plotone{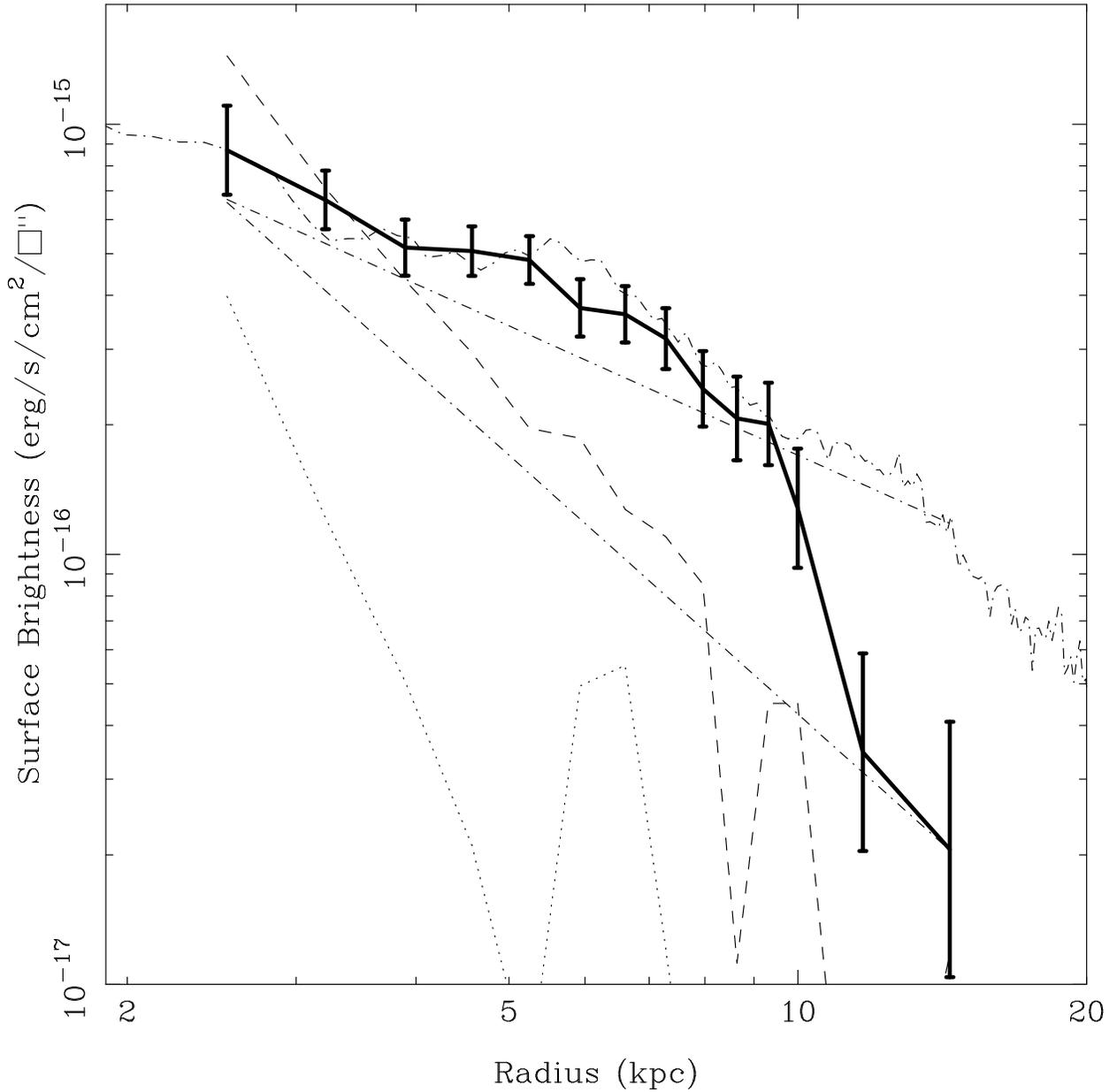}
\caption{Radial surface-brightness profile for the wedge (solid), `anti-wedge' (dashed) and 
continuum (dotted) annuli.  The profile derived from the 
GALEX data for M82 normalized to the 1338 wedge is plotted as a dot-dash line.  The two straight 
dot-dash lines are power-laws of slope -1 and -2.  There is a clear excess of flux in the 
wedge between radii of 3.5 and 10 kpc.  The slope follows the GALEX profile and the 
$\alpha=-1$ power-law well out to 10 kpc.  
At 10 kpc, the wedge profile falls off rather sharply, but the errors are large.
\label{fig:profile}}
\end{figure}

To measure the radial profile of the wedge we have constructed 
semi-circular annuli which cover the same region(s) as the 
angular bins above (Figure~\ref{fig:annuli}).  
We have increased the width of the annuli towards the outer edge of the 
wedge to maintain constant signal-to-noise ratio per annulus.  
The annular photometry again shows that the line emission is 
significantly brighter in the wedge than on either the opposite 
side of the galaxy or in the continuum.  
The surface-brightness profiles for the wedge (solid line), 
`anti-wedge' (dashed line) and continuum (dotted line) 
are shown in Figure~\ref{fig:profile}.  
The errors for each data point were derived by performing identical photometry 
on the corresponding error map.  
There does not seem to be evidence for limb-brightening toward the 
outer edge of the wedge.  Also shown are power-law profiles of the 
form SB$(r) \propto r^{-\alpha}$ with $\alpha$ values 
of 1 and 2 (dot-dash lines).  The wedge profile has a profile much closer to the 
$\alpha = 1$ curve.  
We shall consider these profiles further in the discussion.  

\section{Discussion\label{Discussion}}

The unparalleled spatial resolution provided by {\it HST} and 
the Advanced Camera for Surveys has allowed us to 
observe kpc-scale structures within the radio galaxy 
TN J1338--1942 at $z=4.1$.  The rest-frame ultraviolet continuum and line 
emission of the host galaxy is morphologically complex 
and consists of several distinct components.  
In this section we discuss these features, their implications 
for the formation of this galaxy and attempt 
to construct a possible scenario for the on-going processes in this 
source.

\subsection{On-going Star-Formation}

Are we witnessing the formation of the bulk of the final stellar mass in this radio 
galaxy?  Previous studies based on ground-based optical and millimeter imaging 
suggest that the current star-formation rate (SFR) within TN J1338 is very 
high, of order several hundred solar masses per year \citep*{VKM+02,CARLOSMM}.  
The current study not only confirms this estimate, under different assumptions, 
but also determines the 
spatial gradients in star-formation activity.  This new spatial information 
allows us to constrain both the timescale and possible 
physical origin of the current star-formation episode.  

There are two methods to estimate the SFR in 
TN J1338 using the ACS imaging data; one using the Ly$\alpha$ luminosity and 
the other using the UV continuum luminosity.  
Both estimates are based on the assumption that the UV light we see from 
the galaxy is due solely to star-formation.  
As we have shown earlier, the UV continuum is relatively unpolluted by 
the non-stellar contributors to the aligned light.  
Conversely, for the line emission, the high observed equivalent-widths 
argue in favor of non-stellar excitation \citep*[e.g.,][]{CharlotFall93}.  
Therefore, the line emission provides an upper limit (modulo dust extinction) 
to the SFR, while the UV continuum provides 
a lower limit.  The comparison of these two derived values 
provides an estimate of the mean SFR for each portion of the galaxy.  

For a given Ly$\alpha$ flux we 
calculate the corresponding Balmer line flux and hence the SFR \citep*{Kennicutt98}.  
This SFR is a lower limit locally due to Ly$\alpha$ being a resonantly scattered line and very 
susceptible to dust.  
For example, the Ly$\alpha$ flux of the wedge (region 3 in Figure~\ref{fig:SegMap}) 
is $3.3 \times 10^{-16}$ ergs s$^{-1}$ cm$^{-2}$ corresponding to an intrinsic 
luminosity of $5.2 \times 10^{43}$ ergs s$^{-1}$ at $z=4.1$.  
If we adopt a Case B Ly$\alpha$/H$\alpha$ ratio of $8.7$ \citep*{Brocklehurst71}, 
and use the relation between H$\alpha$ line flux and SFR, we find 
that region 3 is forming stars at a rate of $40 \Msun$  ${\rm yr^{-1}}$.  
The analogous results for the other regions are listed in Table~\ref{stukjetab}.  
The total ``Kron'' Ly$\alpha$ SFR is $290 \Msun$ ${\rm yr^{-1}}$.  
The sum of the individual SFRs does not equal this total; 
underpredicting it by $\approx 90 \Msun$ ${\rm yr^{-1}}$.  
There is a considerable amount of low surface-brightness line emission even 
within the ``Kron'' apeture ($\sim 18 \times 7$ kpc).  We have not corrected 
for the observed HI absorption \citep*{1338SPEC99,Wilmanetal2004abs}.  

The \iband and \zband band probe the rest-frame UV of TN J1338--1942 at
$\sim1500$ \AA~ and $\sim1775$ \AA, respectively.
We can use the ACS continuum magnitudes to estimate the current star
formation rate, under the assumptions
that the UV luminosity is dominated by the light from
late-O/early-B stars on the main sequence.  We measure total (``Kron'') star formation rates of 86 and 96
$M_\odot$ yr$^{-1}$ for the \iband and \zband bands, assuming a 
Salpeter IMF. The derived SFRs are lower limits, 
since they are dependent on the amount and distribution of dust present in
the UV emitting regions. The slope of the UV continuum can also be used to
measure the extinction.  We measure the slope of the continuum from the $\iband-\zband$
color and use a template spectrum of a typical star-forming galaxy
redshifted to $z=4.1$ to convert the measured slope to a color 
excess, $E(B-V)$.  For the template spectrum we have used 
the stellar population synthesis models of 
\citet{bruzualcharlot03} to create a
typical ``Lyman-break'' galaxy spectrum, having an exponentially-declining 
star formation history (with time constant, $\tau=10$ Myr), an age of 70 Myr, 
0.2 $Z_\odot$ metallicity, and a
Salpeter initial mass function.  The parameters of this template are taken from the
best-fit SED at $z\sim3$ of \citet{papovich01}. 
We varied the dust content by applying the
attenuation curve of \citet{calzetti00} to this template. We find
$E(B-V)=0.12$, yielding a dust-corrected SFR of $\sim220$ $M_\odot$
yr$^{-1}$ in good agreement with the emission line estimate above.  
If we repeat this calculation for each discrete region of the galaxy 
(allowing for $E(B-V)$ and SFR to change for each; see Table~\ref{stukjetab}), 
we again find evidence for diffuse UV light and star-formation.  
The sum of the SFRs for all the regions again 
falls short of the total by a factor of 4.5, 
``missing'' $170 \Msun$ yr$^{-1}$.  

\subsubsection{Shocks and Jet-Induced Star-Formation}

One of the primary explanations for the alignment effect 
is that the passage of the radio jet through the 
interstellar gas induces star-formation \citep*[e.g.,][]{Rees89}.  
Strong large-scale shocks associated 
with the expanding radio source 
overpressure molecular gas clouds which then collapse to form stars.  
The presence of powerful shocks in radio galaxies at $z \simlt 2$ 
has been inferred via their 
ultraviolet emission line ratios \citep*[e.g.,][]{BRL00,DeBreucketal00}.  
For TN J1338 most of the important diagnostic emission lines are unobservable from the ground.  
However, we can use morphological information to search for possible signatures of 
shock processes in this radio galaxy.  

In these respects, useful analogies can be drawn 
between TN J1338 at $z=4.1$ and 4C~41.17 at $z=3.8$.  
The shock properties of 4C~41.17 were studied and modeled in detail by 
\citet*{Bicknelletal2000}.  In addition to {\it HST} imaging of 4C~41.17, these authors 
also used deep emission line spectra and high angular-resolution radio 
imaging to study the relationship between the radio source, the gas and the 
stars.  We apply a similar analysis to TN J1338.

A bow-shock is formed at the tip of the 
advancing radio jet (cf. Figure 3 in Bicknell et al. (2000)).  
Due to the unresolved radio structure we cannot robustly determine 
the location of the jet interaction.  We assume that region 4 in the continuum ACS image is 
the primary site where the jet has shocked or is still impacting the gas.  
This continuum knot is very blue and has a morphological structure which is suggestive 
of jet-cloud interaction; namely, a paraboloid oriented along the radio axis.  
From the spatially-resolved optical images 
we estimate the interaction area to be $\sim 2 \times 10^{44}$ cm$^2$ by assuming the emission 
we see is emitted by a spherical shell.  The jet is most likely well-collimated 
and therefore the area of the jet itself is much smaller than the total; we assume 10\%.  
If we assume that most of the momentum flux of the jet is 
dissipated in this interaction, the shock velocity (see Eqn.~1 
of Bicknell et al.) will be greater than: 
\begin{equation}
v_{\rm sh} \simgt 300 - 2000 F_{\rm E, 46}^{1/2} \beta_{\rm jet}^{-1/2} n_{\rm H}^{-1/2} {\rm km\mbox{  }s^{-1}}
\label{eqn:vsh}
\end{equation}
$F_{\rm E, 46}$ is the energy flux of the jet in units of $10^{46}$ ergs s$^{-1}$, 
$\beta_{\rm jet}$ is the relativistic Doppler parameter, and 
$n_{\rm H}$ is the hydrogen density per cm$^{3}$ in the cloud.  
Comparison between model and observed CIV$\lambda\lambda1548,1550$ doublet 
fluxes can help to constrain 
the pre-shock gas density and the energy flux of the jet.  
Assuming that the ACS UV/Ly$\alpha$ image also the 
spatial distribution of CIV, we can constrain the area from which the line 
is being emitted.  We use the observed CIV/Ly$\alpha$ flux ratio 
to convert the Ly$\alpha$ image to a ``CIV'' image.  
If we follow Bicknell et al. (2000) and take $A_p$ to be the projection of the true area of the shock $A_{sh}$ 
in region 4 and predict the CIV line luminosity for our estimated shock velocity 
we find: 
\begin{equation}
{\rm L(CIV)} \approx 2 \times 10^{42} [\frac{\alpha({\rm CIV})}{0.01}]n_{H}(\frac{A_{sh}}{A_{p}})^{-1} {\rm ergs\mbox{  }s^{-1}}
\end{equation}
$\alpha({\rm CIV})$ is the radiative efficiency of the CIV doublet.  
By comparing this to the observed line luminosity ($4.8 \times 10^{42}$ ergs s$^{-1}$) 
we estimate that the pre-shock 
electron number density to be on the order of $n_{\rm H} = 3$ cm$^{-3}$.  

We have estimated the star-formation rate produced by the jet-cloud interaction in 
region 4.  To determine whether these shocks can lead to star-formation rates in excess of 
a couple dozen solar masses per year (for region 4) 
we parameterize the jet-induced star-formation as follows.  
\begin{equation}
{\rm SFR_{jet}} = 26 (\frac{\epsilon}{0.01}) (\frac{f_{\rm gas}}{1.0}) (\frac{\rho}{3 {\rm cm^{3}}}) {\rm A_{sh,44}} v_{\rm sh,1000} \Msun {\rm \mbox{  }yr^{-1}}
\label{eqn:sfrjet}
\end{equation}
We have taken the ambient pre-shock gas number density, $\rho$, to be 3 cm$^{-3}$ as 
inferred above using the line luminosity.  The area of the shock front is taken from 
the image itself and the shock velocity was assumed to be 
between the extremes possible in equation~\ref{eqn:vsh} above.  
The efficiency with which the shocked gas is converted into stars is denoted by $\epsilon$ and the 
gas volume filling factor by $f_{\rm gas}$.  The assumption of order unity filling factor is most 
likely incorrect globally in the galaxy, but more likely to be realistic in 
this limited region where the gas is approximately uniform.  
We conclude that jet-cloud interactions could be responsible for the 
observed star-formation in TN J1338.  

The existing radio imaging data of TN J1338 does not 
have sufficient angular-resolution to make direct correspondence with 
specific features in the ACS images.  However, since the radio structure 
clearly overlaps the optical galaxy the 
comparison with 4C~41.17 and our assumption of a physical connection between 
the radio and optical structures are likely to be justified.

\subsection{The `Wedge'}

The coupling and regulation of star-formation and nuclear activity in forming 
galaxies and their interaction with their environments 
(i.e., ``feedback'') is a key issue in modern astrophysics.  
At low redshift, imaging and spatially-resolved 
spectrocopic observations of starbursting galaxies show that 
supernovae-driven outflows, so-called ``superwinds'', are common among this 
population \citep*[e.g.,][]{HAM90}.  Whether such outflows actually achieve escape velocity and 
enrich the inter-galactic medium (IGM) is still an outstanding question 
\citep*{Heckman2000,Heckmanetal2000,Martin2004}.  
However, these winds, along with nuclear outflows, are the only 
processes observed to transport material into the outer halos 
of galaxies and are therefore prime candidates for injecting the 
metals and energy that are observed in the IGM.  
At high-redshift, where these processes are likely to be even more prevalent due 
to the higher global star-formation rates, the strongest evidence for 
the presence of outflows 
is spectroscopic.  However, the spectroscopic features observed 
are of ambiguous origin and could be due to inflow, outflow or rotation 
(e.g., \citet{vanOjiketal1997} but see also \citet{Adelbergeretal2003}).  
Spatially-resolved imaging of the emission-line gas 
can provide a more certain indication of outflow, if the gas is collimated 
or exhibits the bi-polar morphology of low-redshift superwinds.  
In the next sections we consider several 
possible origins for the wedge: the photoionization cone of an AGN  
or young stellar population, in-situ star-formation, scattering by dust, or 
an ionized outflow associated with a starburst (i.e., a superwind).

\subsubsection{Photoionization}

In several low-redshift Seyfert galaxies cone-shaped regions of high 
ionization are observed, consistent with 
photoexcitation by an active nucleus \citep*[e.g.,][]{WilsonTsvetanov94}.  
In some cases these cones extend to distances of 15-20 kpc 
from the nucleus, similar to the size of the Ly$\alpha$ wedge seen in our ACS image 
\citep*{WilsonTsvetanov94}.  Both the ionized cone and our wedge have 
high equivalent-width.  We derive a lower-limit to the Ly$\alpha$ rest-frame 
equivalent-width of 650\AA\ for the wedge emission.  
Furthermore, powerful radio galaxies are known to photoionize their 
surroundings based on emission line diagnostics and imaging of low-redshift sources.  
Generally both the line and UV continuum emission are elongated and 
aligned with the radio axis, particularly 
at $z>0.7$ \citep*{McCvBSD87,ChambersMileyvanBreugel87}.  

Conversely however, the principal wedge axis is perpendicular to that of the radio source.  
The unified model for AGN posits that the observed radiation 
is anisotropic \citep*[][]{Antonucci93} due to a combination of obscuration 
close to the nucleus and the intrinsically anisotropic radiation.  
For radio-loud galaxies this preferential radiation axis is traced 
by the line connecting the dual radio lobes.  
The misalignment of the wedge therefore argues against 
photoionization due to the AGN or shocks due to the radio jet.
Furthermore, the most likely position of the accreting black hole powering the 
radio emission and therefore also the primary source of hard ionizing radiation is 
where the radio core and $K_S$-band surface-brightness peak coincide.  
The apex of the wedge does not coincide with this position.  
It is possible that a second AGN (this one radio-quiet), coinciding with 
region 4 in Figure~\ref{fig:SegMap}, could ionize the wedge.  
However, this additional black hole would have to have 
its primary axis roughly perpendicular to the radio-loud AGN and 
be much less luminous at $K_S$-band.  While not impossible, this explanation is ad hoc 
and not preferred.  

\subsubsection{In Situ Star Formation or Galaxy Interaction\label{InSituMerger}}

On-going star-formation within the 
wedge itself, and perhaps extending into the outer Ly$\alpha$ halo 
(outside the ``Kron'' radius), would 
also produce bright line emission.  
However, there is a robust upper-limit to the Lyman-$\alpha$ equivalent width 
produced by normal massive 
stars of $W_{\lambda} = 400$ \AA\ \citep*{CharlotFall93}.  For the wedge we estimate 
a significantly higher equivalent width.  We also note that dust extinction 
will decrease the observed equivalent width from its true value; the resonant 
scattering of Ly$\alpha$ photons increases their optical depth relative 
to continuum photons.  This equivalent width argument also applies 
to tidal debris ejected from the galaxy via a merger or interaction.  
It therefore seems unlikely that stars are directly responsible for the 
wedge emission.  

\subsubsection{``Superwind'': Comparison with M82\label{Wind}}

In \citet*{HAM90} the authors use their observations of local starburst galaxies to 
determine the minimal condition for driving a galactic-scale outflow powered by 
supernovae explosions, or a `superwind'.  
They phrase this criterium in terms of the star-formation rate per unit 
area ($\Sigma_{\rm SFR}$), and emprically determine the minimum to be 
$\Sigma_{\rm SFR} \geq 0.1 \Msun$ ${\rm yr^{-1}}$ ${\rm kpc^{-2}}$.  If we adopt this 
minimal value for TN J1338 and apply it to the resolved area where we see 
the wedge emerging from the galaxy (assuming a circular region seen in projection), 
we derive a lower-limit to the SFR over this same area of $1.5 \Msun$ yr$^{-1}$.  
Above we have shown that the SFR for this galaxy greatly exceeds this limit even only 
in region 4 where the wedge may originate.  

Galactic-scale winds have been observed in detail around local starburst galaxies, 
a well-studied example being M82 \citep[e.g.,][]{HAM90}.  
Morphologically, these superwinds are bipolar structures emanating from the galaxy 
nucleus and along the minor axis of the galaxy.  They are detected as emission-line 
filaments, extended X-ray lobes and bipolar thermal dust emission \citep{HAM90}.  
In M82, the emission-line gas is photoionized in the innermost regions and 
primarily shock-excited in the outskirts.  The optically emitting gas 
flows from M82 in filaments which trace the biconical surface.  
The outflows have double-peaked emission lines (due to the two surfaces of the 
cone being separated in velocity) and also, depending on geometry and spectral 
resolution, emission lines with blue-shifted absorption.  
On the basis of these spectroscopic signatures, galactic-winds seem to be a 
generic feature of high-redshift star-forming galaxies \citep*{Pettinietal01}.  
All the existing spectroscopy of TN J1338 has been taken 
along the radio axis where the dynamics of the gas are presumably dominated 
by the AGN outflow and jet-cloud interactions 
\citep*[see above and e.g.,][]{VillarMartinetal99,SolorzanoTadhunterAxon2001}.  
So we must rely on morphology alone to infer the 
presence of a superwind emanating from this radio galaxy along the wedge.  

\begin{figure}
\plottwo{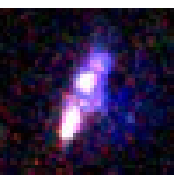}{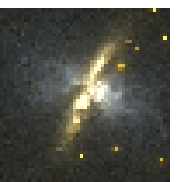}
\caption{left: Color composite of \rband, \iband and \zband ACS images.
right: GALEX near (red) and far-UV (blue) composite image of M82, showing the bipolar outflow.  
In this case the far-UV is interpreted as being continuum light from the starburst that is 
scattered by dust mixed with the outflowing gas \citep*{Hoopesetal}.  
Note the great similarity between the morphologies of the two objects which suggests that 
the wedge emanating from TN J1338 is also a starburst-driven outflow.
\label{fig:GALEX}}
\end{figure}

Comparison of the recent ultraviolet GALEX image of 
M82 \citep*{Hoopesetal} with our TN J1338 image reveals a 
striking degree of similarity (Figure~\ref{fig:GALEX}).  
The scales of the two outflows are somewhat different; in M82 the narrowest 
collimated section is only 1.5 kpc across, while in TN J1338 the similarly defined 
region is approximately twice that.  
The surface-brightness profile is also similar between the wedge and the M82 
far-UV outflow (see Figure~\ref{fig:profile}).  The shallow drop-off of emission-line 
surface-brightness is consistent with having shock ionization dominate 
at these large radii rather than photoionization by a central source.  

Why do we see only one side of the presumably bipolar outflow?
This could be due to obscuration of the line emission 
on one side of the galaxy.  If the radio galaxy 
is flattened (perpendicular to the plane of the sky) and inclined 
with respect to the line-of-sight then any dust in the galaxy would 
naturally obscure the side tilted away from the observer.  
This is the primary explanation for the observed asymmetry in the bipolar 
superwind in M82 \citep*{ShopbellBlandHawthorn1998}.  Alternatively, the lack of line emission 
on the northeast side may be due to a strong gradient in the 
ambient gas density.

Alternatively, if the pressure is higher on the northeast side of the 
galaxy, a situation which could arise due to the motion of the radio 
galaxy within the protocluster medium, i.e., ram pressure, then the outflow 
would be impeded on this side of the galaxy.  
\citet*{MarcoliniBrighentiDErcole04} have made simulations (albeit for 
dwarf galaxies) that show that motion through the IGM 
does not greatly affect the dynamics of galactic outflows until the ram pressure 
becomes comparable to the static thermal pressure of the galactic ISM.  This result 
should be extendable to the case of TN J1338.  To estimate the thermal pressure 
in the galaxy we must first estimate a density for the gas.
If we assume that the observed Ly$\alpha$ emission is due to Case B recombination 
at $T=15000$K, then we can use the fiducial Ly$\alpha$/H$\alpha$ ratio of $\sim 10$  
to deduce the number of ionized hydrogen atoms.  Given the observed geometry 
of the wedge we can assume further that the emitting gas is contained in 
the surface of a cone with half-angle 30$^\deg$ and length 10 kpc.  The thickness 
of the gas layer cannot be larger than a few hundred parsec due to the absence of significant 
limb-brightening in the azimuthal profile of the wedge.  Consequently, the 
number density and mass of ionized hydrogen are:
\begin{equation}
n_e = 1.0 L_{H\beta, 41}^{1/2} V_{\rm cone, kpc^3}^{-1/2} {\rm cm^{-3}}\\
\end{equation}
\begin{equation}
M_{\rm HII} = 7.6\times10^{8} \mu_{p} L_{H\beta, 41}^{1/2} V_{\rm cone, kpc^3}^{1/2} \Msun
\end{equation}
Where $\mu_{p}$ is the mean particle mass, which we have taken to be the proton mass, 
$L_{H\beta, 41}$ is the $H\beta$ luminosity in units of $10^{41}$ ergs s$^{-1}$.  
The true electron density is likely to 
be higher than this, but only within smaller clouds or filaments as is seen in the 
M82 outflow.  The resulting thermal pressure is equivalent to the ram pressure 
produced by a relative velocity of 300 km s$^{-1}$ (if the surrounding gas 
has a density 1/1000th of that in the wedge).  This is some evidence that the 
radio galaxy is not in the center of the galaxy overdensity, and that motion towards the 
center would provide the requisite ram pressure (Intema et al. 2005, in prep.).  
We conclude that the wedge is likely to be a supernovae-driven 
ouflow, with the current episode of star-formation possibly triggered by the radio jet.  This 
superwind is one-sided due to ram pressure inhibiting the flow on one side.

\subsection{The Outer Ly$\alpha$ Halo\label{Disc:OuterHalo}}

\begin{figure}
\epsscale{0.5}
\plotone{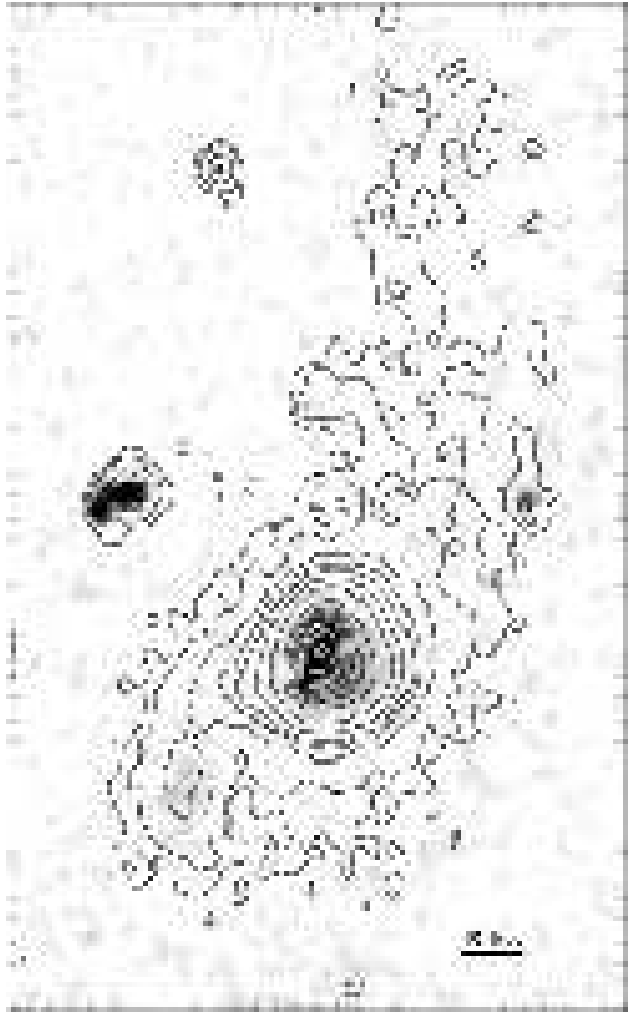}
\caption{Continuum-subtracted ACS $r$-band image overlaid with the VLT 
narrow-band image (contours).  North is up and East to the left.  The fiducial 
bar in the lower-right of the image is 10 kpc long.  
The field-of-view is $15\arcsec \times 22\farcs5$, each major tickmark is one arcsecond.  
Note the correspondence between the wedge and the larger scale structure of the halo 
seen in the ground-based image.  The `filament' extending to the north is aligned with the 
radio axis, although it extends far beyond the northern lobe.
\label{fig:acsvlt}}
\end{figure}

In Figure~\ref{fig:acsvlt}, the extended Ly$\alpha$ (out to $\sim 100$ kpc), 
as detected in the very 
deep VLT narrow-band image, is shown as contours.  The TN J1338 halo is somewhat has an 
asymmetric `plume' which is aligned with the radio axis.  
It is clear from the underlying ACS image showing the wedge 
that there is a natural connection between the 
high surface-brightness wedge (out to 20 -30 kpc) and the 
larger scale lower surface-brightness halo along the southwest direction.  
However, the halo at larger distances appears to be aligned with the radio axis of 
TN J1338.  What is the relation, if any, between the wedge and the large-scale 
Ly$\alpha$ structure?

If the wedge is an outflow, as we conclude above, the resulting 
bubble will stall at some radius where gravity and the amount of swept up intergalactic 
matter balance the input energy.  The gas deposited at this radius will naturally flow 
along the boundary of the excavated cavity and follow any density gradients in the 
ambient medium.  The halo-radio alignment would then be a natural consequence if lower density 
regions were preferentially along the radio axis.  This would be the case if either the 
current radio source extended further out in radius than our current radio observations 
indicate, or that the radio source was previously (either during this same accretion 
episode or during an earlier one) much larger and had excavated the region along the 
radio axis.  However, there is no evidence in the current radio data for a relic 
radio source at larger distances.  
In any case, if the ionized gas has originated in the starburst, the observed 
alignment implies that the AGN had already `imprinted' the region before the starburst 
was triggered.  We discuss this possibility and its implications 
further in the next section.

\subsection{A Self-Consistent Scenario}

The host galaxy of the powerful radio source TN J1338--1942 at $z=4.1$ is 
unique.  It is arguably the youngest brightest cluster galaxy 
known to date, and has been the subject of several multi-wavelength investigations.  
In this paper we have presented imaging from {\it HST}/ACS which reveals several 
interesting morphological and broad-band spectral features in this radio galaxy.  
In this section we attempt to construct a plausible and 
self-consistent story of the past, present and future of TN J1338.  

\begin{figure}
\epsscale{1.0}
\plotone{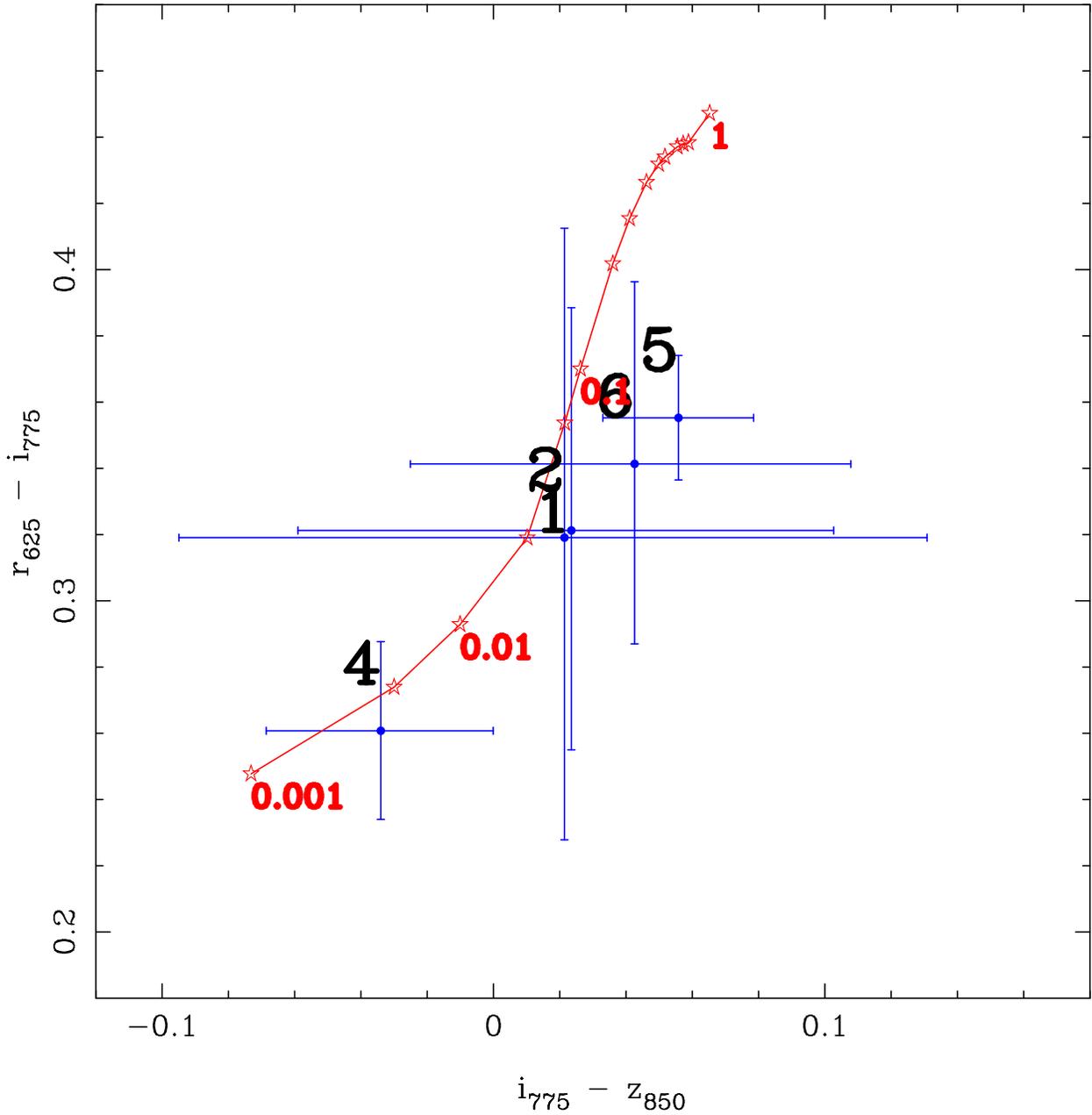}
\caption{\rband - \iband  vs. \iband - \zband color-color diagram for the 
individual regions in TN J1338.  We have excluded the wedge (region 3) from this plot since 
it is a clear outlier, \rband - \iband $ = -0.11$ and \iband - \zband $ = -0.39$.  
The estimated Lyman-$\alpha$ flux has been subtracted from the \rband 
magnitude for each region using a continuum fit to the \iband - \zband color. 
The overplotted red line shows the colors for a constant star-formation model with ages indicated 
in Gyr, $E(B-V)=0.1$ and $Z=0.4 Z_{\odot}$.
\label{fig:ri_iz}}
\end{figure}

The host galaxy of TN J1338 appears to be forming stars at a high rate.  
None of the non-stellar processes known to produce the alignment effect in 
other galaxies can be dominant in this case.  
The ACS data presented here reveals a morphology that is consistent with 
most of the star-formation being triggered by the passage of the 
radio jet.  In Figure~\ref{fig:ri_iz} we show a color-color diagram for the 
discrete regions within the radio galaxy.  The color difference 
between regions 4 and 5 may be an age effect (region 4 is also bluer in $\iband - K_S$).  
We use the overplotted model colors in Fig.~\ref{fig:ri_iz} to derive an age 
difference between the two regions.  The model has constant star-formation with 0.4 solar 
metallicity and $E(B-V) = 0.1$, the ages are labeled at timesteps of 
1, 10, 100 and 1000 Myr.  We estimate from the comparison between the model points and 
the data that the age difference is between 25 and 200 Myr.  
This matches within the (large) errors the shock travel time from the 
radio core (in region 5) to the jet-cloud interaction in region 4 (a distance of 
$\sim 7$ kpc with a 300 km/s shock).  
We do not see very extended star-formation which may be triggered by 
the expanding radio cocoon.  
The energy injection from the AGN may be rather isolated to these few 
nodes along the radio jet itself.  Therefore, we suspect that the 
large-scale Lyman-$\alpha$ gas has an origin apart from the AGN, namely 
in the newly-formed stars.

Once the prodigious star-formation is initiated, the supernovae explosions 
expel the ionized gas into the surrounding media by means of a superwind.  
In addition, ram pressure is stripping the gas from the star-forming 
regions as the radio galaxy moves through the ICM/IGM.  
This gas remains ionized primarily via shock excitation at the interface 
between the outflow and the ambient intergalactic gas.  
At a distance of approximately 20-30 kpc from the nucleus, 
the outflow reaches 
pressure equilibrium with the IGM.  The gas rapidly cools at this 
interface, consistent with the boundary to the 
Ly$\alpha$ halo observed at this region.  
This cooling gas will follow the density gradients 
in the IGM.  The very low surface-brightness line emission seen to the northeast 
of the radio galaxy, which is aligned with the radio axis, is probably 
gas which originated in the outflow, cooled, and is now being re-excited 
by low-luminosity shocks associated with the radio source.  

The observed rest-frame $B$ magnitude of TN J1338 is approximately 1.5-2 magnitudes 
brighter than that of the six brightest cluster galaxies at $z \sim 1$ 
observed with ACS \citep*{PostmanMDR}.  
If we use the same star-formation model as above (constant 
for 1 Gyr and 0.4 solar metallicity) and age the galaxy from 1 Gyr at $z=4.1$ to 
5.2 Gyr at $z=1$ the galaxy fades by 3.3 magnitudes.  This would 
imply that some additional star formation or merging must occur during those 4.2 
Gyrs for TN J1338 to match the luminosity of the $z \sim 1$ BCGs.  
This is certainly not surprising.  It is interesting to note as well, that several of these 
BCGs have a nearby bright companion, of almost equal luminosity ($\Delta M \simlt 0.1$ mag), with which it seems to 
be destined to merge.  The resulting increase in 
luminosity would nearly make up the 1 magnitude 
of ``extra'' fading seen for the 1 Gyr constant star formation model.

It is debatable whether one should attempt to draw conclusions for an entire population 
of sources (either BCGs or radio galaxies in this case) based on observations 
of a single example. TN J1338 may be a galaxy which in a special phase of 
its evolution, or alternatively, it may be a special source whose history cannot be 
generalized to describe other galaxies.  However, studies of ensembles of radio galaxies, including 
their luminosity functions and duty-cycles, suggest that the 
space density of radio source hosts at high-$z$ are roughly in agreement with the 
density of BCGs at low redshift and the density of non-RG overdensities at $z\sim3$ \citep*{West94,VKM+02}
 
\section{Conclusions and Future\label{Conclusions}}

The host galaxy of powerful radio source TN J1338--1942 shows 
signatures of several feedback processes which connect 
the black-hole, stellar host and intergalactic medium.  
The elongated and multi-component ultraviolet continuum is aligned 
with the FRII radio axis and is likely to be due to 
emission from young stars being formed along the jet axis.  Interpretation 
of this light in terms of jet-induced star-formation is 
consistent with the observations.  
There is, however, also evidence for considerable 
star-formation outside the highest surface-brightness regions.
If the current star-formation 
rate has been constant over the jet travel-time from the radio 
core to the site of the presumed jet-cloud interaction, this 
process could have produced in excess of $10^{11} \Msun$ of 
stars.  Data from the {\it Spitzer Space Telescope} 
will allow us to determine the total stellar mass of 
TN J1338 and help verify our hypothetical star-formation history.  

We interpret the ACS wedge of Ly$\alpha$ emission as a superwind 
driven by the winds and supernovae explosions associated with prodigious 
star-formation activity.  This outflow connects 
with the larger-scale Ly$\alpha$ halo.  The initial source of the 
halo gas is then within the starburst, and is possibly enriched.  
Deep spectroscopy from the ground along the wedge axis would 
help us determine the ionization mechanism more definitively.  
In particular, covering the Ly$\alpha$, CIV, HeII and CIII] 
emission-lines may enable us to also measure the enrichment of the 
outflowing gas.  An on-going ACS program using the narrow-band (ramp) 
filter (PI: W. van Breugel) to image several high-redshift radio galaxies 
in Ly$\alpha$ will discover how prevalent such wedge features are 
in this population.

\acknowledgments

We thank M. Seibert, C. Hoopes and the rest of the GALEX team for 
providing their M82 images ahead of publication.  
We gratefully acknowledge M. Lehnert for helpful 
discussions and the anonymous referee for valuable comments.  
ACS was developed under NASA contract NAS 5-32865, and this research 
has been supported by NASA grant NAG5-7697 and 
by an equipment grant from  Sun Microsystems, Inc.  
The {Space Telescope Science Institute} is operated 
by AURA Inc., under NASA contract NAS5-26555.
We are grateful to K.~Anderson, J.~McCann, S.~Busching, A.~Framarini, S.~Barkhouser,
and T.~Allen for their invaluable contributions to the ACS project at JHU.

\clearpage

\begin{deluxetable}{lll}
\tabletypesize{\scriptsize}
\tablewidth{0pt}
\tablecaption{``Kron'' Aperture Photometry\label{tab:apmag}}
\tablehead{
\colhead{Bandpass} & \colhead{AB Magnitude} & \colhead{Estimated Line} \\[0cm]
&&\colhead{Contamination\tablenotemark{a}}}
\startdata
\gband & $25.92 \pm 0.28$ & $\simlt 0.05$ mag\tablenotemark{a}\\
\rband & $22.46 \pm 0.01$ & $\sim 1.3$ mag (Ly$\alpha$)\\
\iband & $23.23 \pm 0.03$ & $\sim 0.3$ mag (CIV$\lambda1549$,HeII$\lambda1640$)\\
\zband & $23.11 \pm 0.04$ & $\sim 0.2$ mag (HeII$\lambda1640$,CIII]$\lambda1909$)\\
$K_S$ & $21.9 \pm 0.2$ & $\simlt 0.01$ mag\tablenotemark{b}\\
\enddata
\tablenotetext{a}{Calculated using the observed spectrum of TN J1338--1942}
\tablenotetext{b}{Calculated assuming the composite spectrum of \citet{McCarthy93}}
\end{deluxetable}

\clearpage

\begin{deluxetable}{ccccccccc}
\tabletypesize{\scriptsize}
\tablewidth{0pt}
\tablecaption{Six Easy Pieces\label{stukjetab}}
\tablehead{
\colhead{Number} & \colhead{Piece} & \multicolumn{3}{c}{AB Magnitude} & \colhead{Lyman-$\alpha$ Flux} & \colhead{Lyman-$\alpha$ Rest-frame} & \colhead{Extinction} & \colhead{Star Formation Rate}\\[0cm]
&& \rband\tablenotemark{a} & \iband & \zband & ($10^{-16}$ ergs/s/cm$^2$) & Equivalent Width (\AA) & $E(B-V)$ & UV Continuum (Ly$\alpha$)\tablenotemark{b}
}
\startdata
1 & Line 1 & 27.14 & 26.82 & 26.80 & 1.17 & 448 & 0.02 &  3 (17)\\
2 & Line 2 & 27.66 & 27.34 & 27.32 & 0.80 & 492 & 0.02 & 2 (12)\\
3 & Wedge & 26.19 & 26.36 & 26.81 & 4.20 & 645 & 0.0 & 3 (61)\\
4 & Continuum 2 & 25.74 & 25.48 & 25.51 & 3.73 & 390 & 0.0 & 9 (54)\\
5 & Continuum 1 & 25.33 & 24.98 & 24.92 & 1.69 & 121 & 0.06 & 24 (24)\\
6 & Shock & 26.54 & 26.20 & 26.15 & 2.45 & 536 & 0.04 & 7 (35)\\
\enddata
\tablenotetext{a}{Calculated for the Ly$\alpha$ subtracted image, where the line flux 
is determined from power-law continuum fits to the \iband and \zband bands for each piece}
\tablenotetext{b}{The star formation rate as calculated by the UV continuum 
magnitude and the Ly$\alpha$ flux (in parenthesis)}
\end{deluxetable}

\clearpage

\clearpage

\clearpage

\clearpage

\clearpage

\clearpage

\clearpage

\clearpage

\clearpage

\clearpage

\clearpage

\clearpage

\end{document}